\documentstyle[12pt,epsfig,amssymb,citesort]{article}

\renewcommand{\arraystretch}{2}

\def\alps{\alpha_s}

\def\msbar{${\rm{\overline{MS}}}$}
\def\alps1{${\cal{O}}(\alpha_s^1)$}
\def\alp{${\cal{O}}(\alpha_s^0)$}
\def\alpsq{${\cal{O}}(\alpha_s^2)$}
\def\oalps{${\cal{O}}(\alpha_s)$}

\begin{document}

\setlength{\baselineskip}{0.75cm}
\setlength{\parskip}{0.45cm}
\begin{titlepage}
\begin{flushright}
DO-TH 00/04 \linebreak
March 2000
\end{flushright}
\vskip 0.8in
\begin{center}
{\Large\bf Fragmentation Functions from  
Flavour-inclusive\\ and Flavour-tagged $e^+e^-$ Annihilations} 
\vskip 0.5in
{\large S.\ Kretzer}
\end{center}
\vskip .3in
\begin{center}
{\large Institut f\"{u}r Physik, Universit\"{a}t Dortmund \\
D-44221 Dortmund, Germany }
\end{center}

\noindent
{\large{\underline{Abstract}}}

\noindent
Fitting $Z^0$-pole data from ALEPH and SLD, and TPC 
data at a lower c.m.s.\ energy, we fix the boundary
condition for NLO parton$\rightarrow$hadron 
(hadron$=\pi^\pm, K^\pm, \sum_h h^\pm$)
fragmentation functions (FFs) at the low resolution scale of the radiative 
parton model of Gl\"{u}ck, Reya and Vogt (GRV). 
Perturbative LO$\leftrightarrow$NLO stability is investigated.  
The emphasis of the fit is on information on the fragmentation process
for individual light ($u,d,s$) and heavy ($c,b$) quark 
flavours where we comment on the factorization scheme for
heavy quarks in $e^+e^-$ annihilations as compared to deep inelastic
production.  
Inasmuch as the light quark input parameters 
are not yet completely pinned down by measurements
we assume power laws to 
implement a physical hierarchy among the FFs
respecting valence enhancement and strangeness suppression
both of which are manifest from recent leading particle measurements.
Through the second Mellin moments of the input functions 
we discuss the energy-momentum sum rule for massless FFs. 
We discuss our results in comparison to previous fits and 
recent 3-jet measurements and formulate present uncertainties 
in our knowledge of the individual FFs.    
\end{titlepage}

\newpage
\section{Introduction}
\label{epemchap}

In this article we consider the production of identified light hadrons 
in $e^+e^-$ collisions or more precisely the production of charged 
hadrons built from light up, down and strange valence quarks, 
i.e.\ dominantly pions, kaons and nucleons. 
Within perturbative QCD the light hadron production dynamics of $e^+ e^-$ 
annihilations inevitably comprise a 
nonperturbative long distance component from
the nonperturbative hadronization of perturbatively produced partons.
We will fix this latter nonperturbative component
in the canonical QCD framework of parton$\rightarrow$hadron fragmentation 
functions (FFs)
at the low resolution input scale of the radiative GRV
parton model \cite{grv98}.
In the naive parton model the FF $D_p^h(z)$ has the interpretation of a 
probability density that some final state parton $p$ hadronizes 
(or {\it{fragments}}) into a mean  number $D_p^h(z) dz$ of hadrons $h$ per 
$dz$ where $z$ is the fractional momentum which $h$ receives from the parton. 
Field and Feynman have constructed \cite{fife} an early set of 
mesonic fragmentation functions footing solely on this intuitive probabilistic 
interpretation. Today's state of the art embeds the $D_p^h(z)$
functions in the framework of QCD factorization 
theory \cite{fact} including explicit operator definitions \cite{cs}.
Fragmentation functions are the final state analogues 
of the initial state parton distribution functions (PDFs) and 
precisely as the PDFs do the FFs parametrize our ignorance of QCD 
bound state dynamics. Fragmentation functions into identified hadrons 
are therefore in their own right an interesting source of information 
on the hadronization process and they are a necessary 
ingredient to interpret present and future measurements of any 
one-particle-inclusive hard cross section within fundamental (perturbative) QCD 
theory as compared to Monte Carlo model approaches. 
Whereas PDFs are mainly determined from fully inclusive DIS 
the cleanest extraction of FFs is from $e^+e^-$ collisions.
FFs pinned down in $e^+e^-$ can, by universality, then
be applied to, e.g., the hadro- or leptoproduction of identified hadrons
and compatibility with the transverse momentum spectrum 
produced in photon-proton collisions was found in \cite{bkk4}.  
From a combined use of
both - fragmentation {\it and} parton distribution
functions - in one hadron inclusive deep inelastic scattering
information on the initial state parton flavour can be obtained from 
leading particle effects where a high energetic hadron inside a jet
remembers the valence parton it has been produced off 
\cite{hermes,emc,geiger}. 
Parton information from 
semi-inclusive measurements based on an understanding  
of the fragmentation
process also extends to polarized DIS where the spin flavour structure
of the nucleon sea is still rather unknown \cite{polsidis,hermes}. 
In this article we will restrict ourselves to the
defining $e^+ e^-$ process and leave other applications 
to future work \cite{forthcoming}. 

The recent years have seen much effort 
\cite{nasweb,rolli,bkk3,bkk4,bkkelse,indu,fsv,aleph,aurenche}
to establish a similar technical skill for FFs as for existing 
sets of PDFs \cite{grv98,ref:mrst,cteq5}. 
We will take a further step towards this goal and for this
purpose we concentrate mainly on data which constrain the  
flavour decomposition
of fragmentation spectra; such data sets exist 
at the $Z^0$ pole
for charged $\pi^\pm$ and $K^\pm$ mesons from SLD \cite{sldpk}
and for the inclusive sum over charged hadrons from ALEPH
\cite{aleph}. Lower energy counterparts from TPC \cite{tpc}
will be included to correctly take QCD scaling violations
- which have been re-established recently in \cite{bkk3,bkk4,aleph} -
into account. 
The flavour separation
confronts us with the question of how to treat heavy flavours 
within $e^+e^-$ production dynamics which we will answer in some detail.    
Especially the SLD data sets on flavour tagged fragmentation into
$\pi^\pm$ and $K^\pm$ mesons contain new flavour information not 
included in previous fits \cite{bkk4} to which we compare our results,
thereby formulating the present uncertainty of our knowledge on the 
individual FFs. 

\section{Parton Fragmentation in $e^+e^-$ Collisions 
Beyond the Leading Order}
\label{NLOsec}

The NLO framework\footnote{The formulae below include the 
LO framework in an obvious way by dropping subleading terms.} 
for one-hadron-inclusive $e^+e^-$ annihilations
is well known since long 
\cite{aemp,nasweb,fsv,fupe,pijff,grvff,swff,bkk3,bkk4} 
and we restrict ourselves to a brief theoretical introduction 
here in
which we closely follow Ref.\ \cite{nasweb} in notation. 

To be specific we will consider the reaction
\begin{equation}
\label{epluseminus}
\frac{d\sigma^{(e^+e^-\rightarrow \gamma,Z^0 \rightarrow h X)}}{dz}
\equiv \frac{d\sigma^h}{dz}
= \frac{d\sigma_T^h}{dz} + \frac{d\sigma_L^h}{dz}\ \ \ , 
\end{equation}
where 
\begin{equation}
z\equiv 2 E_h/Q=2P_h\cdot q/Q^2
\end{equation} 
is the energy\footnote{
In cases where experimental data  are presented in the  
{\it momentum} scaling variable $z_p\equiv 2 p_h/Q$ we will transform the 
data to $z$ using the relativistic energy momentum relation $E_h^2=p_h^2+m_h^2$
and {\it assuming} pion production dominance $(m_h=m_\pi)$ whenever the 
hadron $h$ is
not specified. 
}  
$E_h$ of the observed hadron 
scaled to the beam energy $Q/2\equiv \sqrt{s}/2$;\footnote{
Though $s\equiv Q^2$ we will use both variables
- $s$ and $Q^2$ - in the following 
in their r\^ole as the c.m.s.\ energy and perturbative hard scale, respectively.}  
with the positron/electron beam
momentum $P_{e^\pm}=(Q/2,0,0,\pm Q/2)$ and $q=P_{e^+}+P_{e^-}$.
The r.h.s.\ of Eq.\ (\ref{epluseminus}) distinguishes the contributions
from transverse (T) and longitudinal (L) virtual bosons, where 
the polarization axis is in the direction of the momentum of the observed
hadron $h$. Experimental data are commonly not presented for the absolute 
cross section in Eq.\ (\ref{epluseminus}) but for the normalized
distribution 
\begin{equation}
\frac{1}{N_{tot}}\ \frac{\Delta N^h}{\Delta z} \rightarrow
\frac{1}{\sigma_{tot}}\ \frac{d \sigma^h}{dz}\ \ \ ;\ {\mbox{as}}
\ \Delta z \rightarrow 0\ \ \ ,
\end{equation} 
where $\Delta N^h$ counts the registered $h$-events per bin $\Delta z$
and $N_{tot}$ denotes the inclusive sum of hadronic events.   
Within perturbative QCD and up to \alps1\ the total hadronic cross section 
$\sigma_{tot}$ is given by
\begin{equation}
\label{sigmatot}
\sigma_{tot} = 
\sum_q \sigma_0^q(s) \left[
\left( 1 + \frac{\alpha_s(Q^2)}{\pi}\right)
+ {\cal{O}}\left(\frac{m_q^2}{Q^2}\right) \right]
+ {\cal{O}}\left(\frac{\Lambda_{QCD}^4}{Q^4}\right) 
\end{equation}
where the parton model, i.e.\ \alp , electroweak cross sections 
$\sigma_0^q$ for producing a $q{\bar q}$ pair
are given in the Appendix.
The power suppressed terms arise from perturbative quark mass 
effects [${\cal{O}}(m_q^2/Q^2)$]
or nonperturbative higher operator matrix elements 
[${\cal{O}}(\Lambda_{QCD}^4/Q^4)$]. 
We will discuss 
the quark mass effects in Section \ref{heavy} 
where we give
ample reasons for our choice of the factorization scheme. Higher operators are 
not considered in this article.

QCD factorization theory predicts that
\renewcommand{\arraystretch}{1}
\begin{eqnarray}    
\label{fragconv}
\frac{d\sigma_{P=T,L}^h}{dz} &=&  \ \sum_{i=\left\{ 
{\tiny{
\begin{array}{c} 
q=u,d,s,... \\ 
{\bar q} = {\bar u},{\bar d},{\bar s},... \\
g
\end{array}}}\right. 
}\ \Bigg[\ 
\int_z^1 \frac{d\zeta}{\zeta} C_P^i \left(\zeta,Q^2,\mu^2_{F,R}\right) 
D_i^h\left(\frac{z}{\zeta}
,\mu^2_F\right)  \\ \nonumber 
&+& \left. 
{\cal{O}}\left(\frac{m_q^2}{Q^2}\right) \right]
+ {\cal{O}}\left(\frac{\Lambda_{QCD}^n}{Q^n}\right)
\end{eqnarray}
\renewcommand{\arraystretch}{1.3}
where the power $n$ of the nonperturbative corrections \cite{power} 
cannot be determined from an operator product expansion based analysis, 
contrary to $n=4$ in
Eq.\ (\ref{sigmatot}) and to deep inelastic scattering where higher
twists are known to be suppressed by a power $n=2$.  
As anywhere $\mu^2_{F,R}$ are the factorization and renormalization 
scale, respectively, and we will set them both equal to 
$\mu^2\equiv\mu^2_{F,R}=Q^2$
in the applications. The treatment
of charm and bottom in the sum over quark flavours $q=u,d,s,...$ in (\ref{fragconv})
will be discussed in Section \ref{heavy}. 
The coefficient functions
$C_P^i$ are given up to \alps1\ in the \msbar\ scheme \cite{aemp,fupe,nasweb} by
\begin{eqnarray} \nonumber
\label{epemcoeffs}
C_T^q \left(\zeta,Q^2,\mu^2_{F,R}\right)
&=& \left[ \delta (1-\zeta ) + \frac{\alpha_s(\mu^2_R)}{2 \pi}\ C_F\ c_T^q\left(\zeta,
\frac{Q^2}{\mu^2_F}\right)\right]\sigma_0^q (s) \\ \nonumber
C_T^g \left(\zeta,Q^2,\mu^2_{F,R}\right)
&=&  \frac{\alpha_s(\mu^2_R)}{2 \pi}\ C_F\ c_T^g\left(\zeta,
\frac{Q^2}{\mu^2_F}\right) \sum_q \sigma_0^q (s) \\ \nonumber
C_L^q \left(\zeta,Q^2,\mu^2_{F,R}\right)
&=&  \frac{\alpha_s(\mu^2_R)}{2 \pi}\ C_F\ c_L^q\left(\zeta,
\frac{Q^2}{\mu^2_F}\right) \sigma_0^q (s) \\
C_L^g \left(\zeta,Q^2,\mu^2_{F,R}\right)
&=&  \frac{\alpha_s(\mu^2_R)}{2 \pi}\ C_F\ c_L^g\left(\zeta,
\frac{Q^2}{\mu^2_F}\right) \sum_q \sigma_0^q (s) 
\end{eqnarray} 
and for antiquarks $C_{T,L}^{\bar q}=C_{T,L}^{q}$. 
The $c_{T,L}^{q,g}$ and $\sigma_0^q$ can be found in the Appendix. 
The \alpsq\ contributions to the coefficient functions $C_{T,L}^{q,g}$ in 
(\ref{epemcoeffs}) are known \cite{rijne} but are of next-to-next-to-leading 
order and will therefore not be considered in this NLO analysis, with
one exception to be discussed in Section \ref{sigl}.   

In the \msbar\ scheme the parton model expectation
\begin{equation}
\label{esum}
\int_0^1 dz\ z\ \sum_h\ D_i^h(z) = 1
\end{equation}
that the entire parton energy is shared by the parton's fragmentation 
products is preserved under renormalization 
[$D_i^h(z) \rightarrow D_i^h(z,\mu^2)$] of the fragmentation functions
by energy conservation 
\begin{equation}
\label{conserve}
\frac{1}{2 \sigma_{tot}}
\ \int_0^1 dz\ z\ \sum_h\ \frac{d\sigma^h}{d z}=1
\end{equation}
where $d \sigma^h$ is related to the $D_i^h$ via Eqs.\ (\ref{epluseminus}),
(\ref{fragconv}).
The renormalized
fragmentation functions $D_i^h$ obey massless Altarelli-Parisi-type
renormalization group equations
\begin{equation}
\label{apff}
\frac{\partial D_j^h(z,Q^2)}{\partial \ln Q^2} =
\sum_i\ \int_z^1 \frac{d\zeta}{\zeta}
\ P_{ij}\left(\frac{z}{\zeta},Q^2\right) D_i^h(\zeta,Q^2)
\end{equation}
where the $P_{ij}$ have a perturbative expansion
\begin{equation}
\label{pijff}
P_{ij}(z,Q^2)=\frac{\alpha_s(Q^2)}{2\pi}\ P_{ij}^{(0)}(z)
\ +\ \left(\frac{\alpha_s(Q^2)}{2\pi}\right)^2\ P_{ij}^{(1)}(z) 
\ +\ {\cal O}\left(\alpha_s^3\right)\ \ \ .
\end{equation}
The NLO pieces $P_{ij}^{(1)}$ of
the timelike $P_{ij}$ in (\ref{pijff}) 
differ from their spacelike counterparts. 
The $P_{ij}$ are implicitly understood to represent the 
{\it timelike} splitting functions in 
\cite{pijff}\footnote{A misprint in the second Ref.\ of \cite{pijff} was 
corrected in \cite{grvff,swff}.}. 

The evolution equations will be solved analytically in Mellin
$n$-space as described, e.g.\ in \cite{fupe,grvff}. 
We have included the 
$n$-space expressions \cite{fupe,grvff} for the $c_{L,T}^{q,g}$ 
in Eq.\ (\ref{epemcoeffs}) in 
the Appendix. The timelike
splitting functions (\ref{pijff}) up
to two loop order \cite{pijff} have been transformed
to $n$-space in \cite{grvff}. To keep this article compact  
we do not reproduce these lengthy formulae
nor do we repeat the solution of Eq.\ (\ref{apff}) in
Mellin space. Suffice to say here 
that some functional input forms
for the $D_i^h(z,\mu^2)$ are required where we make the Ansatz
\begin{equation}
\label{ffansatz}
D_i^h(z,\mu_0^2) = N_i\ z^{\alpha_i^h}\ (1-z)^{\beta_i^h}
\end{equation} 
which we assume to hold for light partons ($i=g,u,d,s$ and corresponding 
antiquarks) at the low input scale $\mu^2_0=0.4\ {\rm GeV}^2$ 
of the recent NLO revision \cite{grv98} of the radiative parton model 
\cite{grv90,grv92,grv93,grv94}. 
(Accordingly, $\mu^2_0=0.26\ {\rm GeV}^2$ \cite{grv98}
will serve as the input scale for the LO fits to be discussed 
below.)
As already mentioned, the treatment of heavy
flavours will be specified in the next Section. Along with the low
input scale $\mu_0^2$ we also adopt the 
evaluation of $\alpha_s^{NLO}$ used in \cite{grv98}, i.e.\ we numerically
solve the renormalization group equation
\begin{equation}
\frac{d \alpha_s (Q^2)}{d \ln Q^2} = - \frac{\beta_0}{4\pi}
\ \alpha_s^2(Q^2) -\frac{\beta_1}{16 \pi^2}\ \alpha_s^3(Q^2)
\end{equation}
exactly
by finding the root of
\begin{equation}
\label{root}
\ln \frac{Q^2}{\Lambda^2_f} = \frac{4\pi}{\beta_0 \alpha_s(Q^2)}
-\frac{\beta_1}{\beta_0^2}\ln\left[\frac{4\pi}{\beta_0\alpha_s(Q^2)}
+\frac{\beta_1}{\beta_0^2}\right] 
\end{equation}
with $\beta_0=11-2f/3$ and $\beta_1=102-38 f/3$ and where the number $f$
of active flavours in the quark loop contributions to the beta function
is
\begin{equation}
\label{active}
f={\scriptsize
  \left\{ \begin{array}{l} 3, \mu^2_0<Q^2<m_c^2\\
                           4, m_c^2<Q^2<m_b^2\\
                           5, m_b^2<Q^2<m_t^2  \\
                           6, m_t^2<Q^2
\end{array} \right. }
\end{equation}
with \cite{grv98} $m_{c,b,t}=1.4,\ 4.5,\ 175\ {\rm GeV}$. Eq.\ (\ref{active})
guarantees the continuity of $\alpha_s$ at the transition scales $m_{c,b,t}$
in the \msbar\ scheme up to two loops \cite{ct}, if we furthermore 
adopt \cite{grv98} $\Lambda_{3,4,5,6}=299.4,\ 246,\ 167.7,\ 67.8\ {\rm MeV}$
in NLO (and $\Lambda_{3,4,5,6}=204,\ 175,\ 132,\ 66.5\ {\rm MeV}$ in LO). 
The continuity of $\alpha_s$ at the transition scales $Q_0$, 
where $f\rightarrow f+1$, is guaranteed for $Q_0^2=m_{c,b,t}^2$ \cite{ct} 
from the ratio of the renormalization constants $Z_3$
in a renormalization scheme with $f$ and $f+1$ active flavours, respectively.
Hence the continuity is not affected by the choice of solving
the NLO renormalization group equation exactly as implied by  Eq.\ (\ref{root})
or by an analytical approximation up to some inverse power of $\ln Q^2$, as, e.g., 
in Eq.~(9.5a) of \cite{pdg98}. The exact numerical solution of Eq.~(\ref{root})
is more appropriate \cite{grv98} in the low $Q^2\lesssim m_c^2$ regime and 
will be used over the entire $Q^2$ range.  

\section{The Longitudinal Cross Section $d\sigma_L^h/dz$}
\label{sigl}

We shall comment here briefly on the counting of perturbative orders for the
longitudinal structure function on the r.h.s.\ of (\ref{epluseminus}).
Experimentally $d\sigma_L^h/dz$ is extracted \cite{aleph}
by reweighting inclusive 
hadronic events according to their polar angle ($\theta$) distribution as  
\begin{equation}
\label{reweight}
\frac{d\sigma_L^h}{dz}=\int_{-v}^v d\cos\theta\ \frac{d^2 \sigma^h}{dz d\cos \theta}
\ W_L(\cos\theta ,v)\ \ \ ,
\end{equation}
where the projector $W_L$ was introduced in \cite{nasweb} and
$v=|\cos\theta|_{\max}$ is set by the detector geometry, 
e.g.\ $v=0.94$ for ALEPH \cite{aleph}.

From the theoretical side we can read off Eqs.\ (\ref{fragconv}), 
(\ref{epemcoeffs}) that $d\sigma_L^h/dz$ receives its leading {\it nonzero} 
(finite and scheme-independent)
contribution at \alps1. When considering data on $d\sigma_L^h$ obtained 
from reweighted events as in (\ref{reweight}) we will therefore include
for our {\it next-to-leading order} analysis the \alpsq\ contributions of 
Ref.\ \cite{rijne}. We will, however, treat the \alps1\ coefficients 
$C_L^{q,g}$ in (\ref{epemcoeffs}) as {\it subleading} (NLO) contributions 
to the {\it total} $d \sigma^h=d \sigma_T^h + d \sigma_L^h$ in 
Eq.\ (\ref{epluseminus}) which receives its {\it leading} contribution
from the $d\sigma_T^h$ component to \alp. 

To \alpsq\ the convolutions in (\ref{fragconv}) are most conveniently 
decomposed according to the flavour group
as described in \cite{rijne}. The corresponding coefficient functions
for $d\sigma_L^h/dz$  
are given in Eq.\ (17), (18) and (20) in the second Ref.\ of \cite{rijne}.
For the transformation to Mellin $n$ space 
the nontrivial ones of the required identities can be found
in \cite{grvg,dedu,bk}.

\section{Treatment of Heavy Flavours}
\label{heavy}

Heavy quark initiated jets provide a substantial contribution 
to one-hadron-inclusive $e^+ e^-$ annihilation spectra.
In principle, 
the hard scattering production
of heavy quarks has a well defined and unique perturbative expansion 
within QCD but the residual freedom of arranging the perturbation series 
at finite order leads into scheme ambiguities \cite{schemes}.  
The key question is 
whether the convergence of perturbative calculations  
improves if fixed order calculations are supplied with an additional 
resummation of quasi-collinear logarithms 
$[(\alpha_s/2\pi ) \ln (Q^2/m^2)]^n$
to all orders $n$, where $m$ is the heavy quark mass.
The explicit all order resummation is equivalent
to solving massless evolution equations from an input scale of ${\cal{O}}(m)$.
The boundary condition at the input scale is again calculable at fixed
order and has been derived for $e^+e^-$ annihilations in \cite{melnas}
to \oalps .  

Fixed NLO calculations seem at present to be  
reliable \cite{grs,grv98} for the deep inelastic production of charm and bottom
- dominantly in scattering events off wee gluons at low partonic c.m.s.\ 
energy where 
mass effects are most important \cite{grs}.
Contrarily, in $e^+ e^-$ annihilations
each quark of a primary heavy quark pair 
receives the energy $Q/2\gg m$ at the intermediate boson decay vertex. 
Such a highly relativistic quark will obviously not `feel' its mass anymore and
at the energies which we will consider,
primary up- and down-type quarks\footnote{
We do, of course, not consider the superheavy top quark here.}
are each produced in equal number, 
irrespective of the quark mass.   
We therefore expect heavy quarks in $e^+ e^-$ to behave
essentially like massless partons with mass 
(quasi-)singularities to be resummed. 
In \cite{cagre,naol,nasol99}  
all order NLO collinear resummations were demonstrated 
to be required and adequate 
in order to describe the energy distribution of charm and bottom quarks 
over the partonic final state of high energy $e^+ e^-$ annihilations. 
In particular, all order 
massless resummations for charm quarks were proven to be necessary to 
describe the amount of secondary charm quarks from $g\rightarrow c{\bar c}$
splittings in the parton showering, 
visible in a rise of the cross section at lower $z$ as observed
by OPAL \cite{opal} and ALEPH \cite{hansper} and as opposed to the fixed 
\alpsq\ order contribution $Z^0\rightarrow q {\bar q} g \rightarrow
q {\bar q} c {\bar c}$ \cite{bernetal} 
which turns out to be far too small \cite{naol,nasol99}.
Bottom production has an evolution length which is shorter by the amount of 
$\Delta \ln Q^2=\ln (m_b^2/m_c^2)$ which  
suffices to suppress secondary bottom pairs as 
experimentally observed \cite{sld} in a flat $z\rightarrow 0$ B-spectrum.    

Hence, we will include $q_{{}\atop{\tiny{H}}}\rightarrow h$ FFs mixing
under evolution with their
light parton analogues, i.e.\ describing all long distance 
(collinear parton showering {\it and} hadronization) effects 
of the fragmentation process. Heavy flavours are treated above their
respective \msbar\ `thresholds' $Q_0=m_{c,b}$
as active flavours in the evolution of 
the $D_{q,g}^h(z,\mu^2)$ in Eq.~(\ref{apff}).
We adopt the functional form of Eq.\ (\ref{ffansatz}) also
for $D_{c,b}^h$, i.e.
\begin{equation}
\label{hqffansatz}
D_{i=c,b}^h(z,\mu_i^2) = N_i\ z^{\alpha_i^h}\ (1-z)^{\beta_i^h}\ \ \ ,
\end{equation}
to hold at $\mu_i^2=m_i^2$ along Eq.\ (\ref{active}) which 
guarantees the continuity of $\alpha_s$ at $\mu_i^2$ in NLO. 
The $D_{i=c,b}^h$ are then {\it dis}continuously 
`switched on' in the evolution at $m_i^2$ but enter the 
cross section in Eq.\ (\ref{epluseminus}) only above the partonic 
threshold $Q^2>4 m_i^2$.\footnote{
The partonic threshold $4 m_i^2$ - or similar low resonance threshold
scales \cite{bkk4} - may therefore be considered a natural choice
for $\mu_i^2$. We follow existing PDF sets which 
avoid the discontinuities in $\alpha_s$ induced by $\mu_i \neq m_i$.}
It is certainly questionable \cite{bkk3} to fit pure
QCD fragmentation functions for heavy quarks to inclusive or
heavy quark tagged data because charm and bottom jets are highly 
contaminated by weak decay channels. Adopting such a procedure anyway
is a necessity as long as data corrected for weak decays do not exist.

\section{Fitting Procedure}
\label{fitprod}

Early model attempts \cite{sakai} to generate fragmentation functions entirely 
by QCD dynamics from a delta peak input $D_p^h(z)\propto \delta (1-z)$
are outruled by modern high statistics data \cite{aleph,sldpk,tpc} which 
require smooth 
input functions (\ref{ffansatz}), (\ref{hqffansatz}) to be fitted
to experiment. 
Recently, fragmentation functions have been fitted
to $e^+e^-$ production data using either free fit Ans\"{a}tze
\cite{bkk4} or Ans\"{a}tze strongly constrained by ${\rm SU(3)_f}$ symmetry 
\cite{indu}. We will take an intermediate path and 
constrain free fit Ans\"{a}tze by making power law assumptions 
about valence enhancement and strangeness suppression
within the radiative parton model of Refs.\ \cite{grv90,grv92,grv93,grv94,grv98}. 
The model is originally tailored for parton distribution functions to 
perturbatively generate the high population of quarks and gluons in hadrons at 
low Bjorken $x$. 
It should be clear from the scratch that its predictivity  
cannot be transfered to FFs where a well defined
{\it low} $z$ analogue of the deep inelastic {\it low} $x$ regime is missing
\cite{fsv}.
    
Our aim is to pin down fragmentation functions $D_i^h(z,\mu^2)$
for (anti-)quarks and gluons hadronizing into 
charged pions ($h=\pi^+,  \pi^-\equiv \pi^{+,-}$), 
charged kaons ($h=K^+, K^-\equiv K^{+,-}$)  
and the inclusive sum over charged hadrons [
$\sum_h (h^+ +h^-)\equiv   \sum h^\pm$];\footnote{
Similarly, $\pi^\pm$ will denote $\pi^+ + \pi^-$ to be
distinguished from $\pi^{+,-}\equiv \pi^+, \pi^-$; for kaons accordingly.}
basically from high statistics data at the $Z^0$ peak measured by
ALEPH \cite{aleph} at LEP ($\sum h^\pm$) and by SLD \cite{sldpk} 
($\pi^\pm,K^\pm$) at SLAC. Besides their high statistical accuracy,
these data sets have the advantage of furnishing along with their fully inclusive 
measurements also flavour enriched events which can be used to fix the flavour
structure of the $D_i^h(z,\mu^2)$ to some extent. 
Both data sets distinguish between light quark ($u,d,s$), 
charm, and bottom events where the quark flavour refers to the 
primary $Z^0$-boson decay vertex. Since pure flavour separated sets 
cannot be obtained directly \cite{aleph,sldpk}, Monte Carlo simulations are
required to estimate the flavour composition of the tagged data sets.
From these Monte Carlo studies the flavour enriched event samples  
have been unfolded at a given systematic uncertainty to pure flavour sets 
for the SLD data whereas ALEPH quotes percentages
for each flavour contribution to any of its flavour enriched samples. 
In the latter case some uncertainty will reside in the 
translation of the Monte Carlo studies to the perturbative calculations we
are performing here which can unfortunately not be quantified 
since no systematic errors are quoted \cite{aleph} for the percentages.
Anyway, we will reweight the electroweak couplings in Table \ref{couptab}
in the Appendix
to reproduce the ALEPH flavour composition as quoted in \cite{aleph}.
In order to have our FFs respect QCD scaling violations properly
we include lower scale ($\sqrt{s}=29\ \rm{GeV}$) TPC data \cite{tpc} in our fits
which also furnish flavour information
(unfolded to pure $\{u,d,s\}$, $c$ and $b$ event samples) for $\sum h^\pm$ and
for $\pi^\pm$ but only inclusive measurements for $K^\pm$.  
Needless to say, the fits will be dominated statistically
by the $Z^0$-pole measurements of ALEPH and SLD.     

Our Ans\"{a}tze for the fragmentation functions will be the ones of 
Eqs.\ (\ref{ffansatz}), (\ref{hqffansatz}) for light parton and heavy quark 
fragmentation, respectively, where we assume the following 
symmetries and hierarchy
\begin{eqnarray}
\label{ffsym1}
D_q^{h^+,h^-}=D_{\bar q}^{h^-,h^+};\ \   h=\pi, K 
\\ \label{ffsym3}
D_{d}^{\pi^+}=
D_{s, {\bar s}}^{\pi^+} < D_u^{\pi^+}=D_{\bar d}^{\pi^+}
\\ \label{ffsym4}
D_{\bar u}^{K^+}=
D_{d, {\bar d}}^{K^+}<D_u^{K^+}<D_{\bar s}^{K^+} 
\ \ ,
\end{eqnarray}
where Eq.\ (\ref{ffsym1}) respects charge conjugation and 
Eqs.\ (\ref{ffsym3}), (\ref{ffsym4}) should hold from the valence 
structure of pions and kaons and strangeness suppression. 
The equality in Eq.\ (\ref{ffsym3}) 
seems to be confirmed by `leading particle'
measurements \cite{lpeff} which also indicate 
the second {\it in}equality in Eq.\ (\ref{ffsym4}) at large $z$ 
from the suppression
of secondary $s {\bar s}$ formation which is required to form a $K^+$ from
a $u$ but not from an ${\bar s}$ quark. 
Beyond strangeness suppression,
in $\pi^{+,-}$ or $K^{+,-}$
production we expect differences from favoured valence-type 
(e.g. $u\rightarrow \pi^+$) and unfavoured sea-type (e.g. $s\rightarrow \pi^+$)
fragmentation channels. 
Nevertheless, we assume a universal $z\rightarrow 0$ 
behaviour, determined by the input parameters 
$N_i$ and $\alpha_i$ in (\ref{ffansatz}), 
This assumption is guided by
the idea that the valence enhancement of, say, $u\rightarrow \pi^+$
fragmentation should manifest itself mainly as a 
`leading particle effect' \cite{lpeff}, 
parametrized by $\beta_i$ in (\ref{ffansatz}) as $z\rightarrow 1$.
We will {\it assume} \cite{reyarep} 
\begin{equation}
\label{ffsym6}
\beta_d^{\pi^+}=
\beta_{s, {\bar s}}^{\pi^+}=\beta_{u,{\bar d}}^{\pi^+}
+1\ \ ,\ \ \beta_{\bar u}^{K^+}=
\beta_{d,{\bar d}}^{K^+}
=\beta_{u}^{K^+}+1 = \beta_{{\bar s}}^{K^+}+2
\ \ \ ,
\end{equation}
which suppresses $s {\bar s}$ formation as well as 
sea-type fragmentation as $z\rightarrow 1$.
A linear suppression factor $D_d^{\pi^+}/D_u^{\pi^+}=(1-z)$ 
for sea-type fragmentation is 
compatible with semi-inclusive deep inelastic measurements
\cite{hermes,geiger} in the range $0.1 < z \lesssim 0.8$. These
measurements seem to prefer a factor $\sim (c-z)$ 
with $c\simeq 0.25$ for $z>0.8$. The very large $z$ behaviour
of the $D_i^h(z)$ is, however, not very well constrained by 
$e^+e^-$ measurements and from the theoretical side
soft gluon resummations may become necessary \cite{aurenche}. 
Also, the universality, i.e.\ process-independence of the FFs
in $e^+e^-$ annihilations and semi-inclusive DIS\cite{forthcoming}, 
respectively,  
is not yet well settled at large $z$ \cite{geiger}. 
We therefore keep $D_d^{\pi^+}/D_u^{\pi^+}=(1-z)$ for 
the time being. Our assumptions in Eq.\ (\ref{ffsym6}) are compatible with 
$D_d^{\pi^+}(n=2)\simeq 0.6\ D_u^{\pi^+}(n=2)$ 
where the second moments are defined below in Eq.\ (\ref{mellin}) and
where their ratio can be estimated from
EMC one-pion-inclusive data \cite{emc}.
These data have been analyzed in a parton model context and the 
extracted scale-independent FFs of \cite{emc} can therefore not be 
compared to our fit in more detail. 
The corresponding one-kaon-inclusive measurements in \cite{emc} seem to 
prefer an even stronger suppression of sea-type fragmentation in kaon
formation. For the time being, we do, however, assume the suppression
of sea-type channels to be a universal phenomenon modeled by one extra
power in $(1-z)$ for the input FFs.   
This assumption, as any of our above assumptions on the light flavour 
structure of annihilation data, 
must obviously be expected to be violated to some extent 
if dedicated measurements will become possible. 

If we furthermore assume for simplicity 
\begin{equation}
\label{hqsym}
D_{c,b}^{h^+}=D_{c,b}^{h^-}
\end{equation}
to hold in the heavy quark sector\footnote{
We would expect Eq.\ (\ref{hqsym}) to hold exactly if heavy quark jets were
not contaminated by weak decays.}, 
we are left with the following set of independent input 
parametrizations for pions and kaons:
\begin{eqnarray} 
\label{pansatz}
D_{u,{\bar d}}^{\pi^+}(z,\mu_{0}^2) &=& N_u^{\pi^+} z^{\alpha_u^{\pi^+}}
(1-z)^{\beta_u^{\pi^+}}  \\ \nonumber
D_{s,{\bar s}}^{\pi^+}(z,\mu_{0}^2) &=& 
N_u^{\pi^+} z^{\alpha_u^{\pi^+}}
(1-z)^{\beta_u^{\pi^+}+1} \\ \nonumber
D_{g}^{\pi^+}(z,\mu_{0}^2) &=& 
N_g^{\pi^+} z^{\alpha_g^{\pi^+}}
(1-z)^{\beta_g^{\pi^+}} \\ \nonumber
D_{c,{\bar c}}^{\pi^+}(z,m_c^2) &=& 
N_c^{\pi^+} z^{\alpha_c^{\pi^+}}
(1-z)^{\beta_c^{\pi^+}} \\ \nonumber
D_{b, {\bar b}}^{\pi^+}(z,m_b^2) &=& 
N_b^{\pi^+} z^{\alpha_b^{\pi^+}}
(1-z)^{\beta_b^{\pi^+}} \\ \nonumber \\
\label{kansatz}
D_{{\bar s}}^{K^+}(z,\mu_{0}^2) &=& N_{\bar s}^{K^+} 
z^{\alpha_{\bar s}^{K^+}}
(1-z)^{\beta_{\bar s}^{K^+}}  \\ \nonumber
D_{u}^{K^+}(z,\mu_{0}^2) &=& 
N_{{\bar s}}^{K^+} z^{\alpha_{{\bar s}}^{K^+}}
(1-z)^{\beta_{{\bar s}}^{K^+}+1} \\ \nonumber
D_{d, {\bar d}}^{K^+}(z,\mu_{0}^2) &=& 
N_{{\bar s}}^{K^+} z^{\alpha_{{\bar s}}^{K^+}}
(1-z)^{\beta_{{\bar s}}^{K^+}+2} \\ \nonumber
D_{g}^{K^+}(z,\mu_{0}^2) &=& 
N_g^{K^+} z^{\alpha_g^{K^+}}
(1-z)^{\beta_g^{K^+}} \\ \nonumber
D_{c, {\bar c}}^{K^+}(z,m_c^2) &=& 
N_c^{K^+} z^{\alpha_c^{K^+}}
(1-z)^{\beta_c^{K^+}} \\ \nonumber
D_{b, {\bar b}}^{K^+}(z,m_b^2) &=& 
N_b^{K^+} z^{\alpha_b^{K^+}}
(1-z)^{\beta_b^{K^+}}\ \ \ .
\end{eqnarray}
In practice, we will express 
the normalizations $N_i^h$ through the 
physically more interesting
contributions to the energy integral in Eq.\ (\ref{esum}),
given by the second Mellin moment 
\begin{equation}
\label{mellin}
D_i^h(n=2,\mu^2)\equiv \int_0^1 dz\ z D_i^h(z,\mu^2)
\end{equation}
which would be an exact invariant under evolution in $\mu^2$ only if 
$D_i^h(n=2,\mu^2)$ were the same for all $i$ -- which is not the case.
Anyway, guided by the idea of having an intermediate gluon in the parton 
cascade connecting a valence- and a sea-type quark   
we will reduce the parameter space
$\{N_i^{\pi^{+,-},K^{+,-}}$, $\alpha_i^{\pi^{+,-}, K^{+,-}}$,
$\beta_i^{\pi^{+,-},K^{+,-}}\}$ somewhat by demanding\footnote{
We do not include $D_{{\bar s}}^{K^+}$ in the second average of
Eq.\ (\ref{parcon}) because
gluon fragmentation into a $K^+$ must be strangeness-suppressed.} 
\begin{equation}
\label{parcon}
\begin{array}{c}
D_g^{\pi^+}(n=2,\mu^2_0)= \frac{1}{2}
\left[D_s^{\pi^+}(n=2,\mu^2_0)+D_u^{\pi^+}(n=2,\mu^2_0)\right]
\\ \nonumber
D_g^{K^+}(n=2,\mu^2_0)=\frac{1}{2}
\left[D_d^{K^+}(n=2,\mu^2_0)+D_u^{K^+}(n=2,\mu^2_0)\right]
\end{array}
\end{equation}
which reduces the strength of the evolution of the $D_i^h(n=2,\mu^2)$
considerably and does not influence the quality of the fits.

Samples of inclusive charged hadrons $\sum h^\pm$ are dominated by 
charged pions and adding a kaon background is a reasonable 
approximation for most applications: 
\begin{equation}
\label{hpmapprox}
D_i^{{\tiny \Sigma} h^\pm}\simeq D_i^{\pi^\pm+K^\pm}\ \ \ .
\end{equation}
Eq.\ (\ref{hpmapprox}) is, however, not adequate to compare with
high statistics measurement from the $Z^0$-pole \cite{aleph},
which are sensitive at the ${\cal{O}}(1\% )$-level. 
We will fit a small residue from higher mesons and baryons along
\begin{equation}
\label{residual}
D_i^{{\tiny \Sigma} h^\pm}= D_i^{\pi^\pm+K^\pm}+D_i^{\rm res.}\ \ \ ,
\end{equation}
using again simple Ans\"{a}tze as in (\ref{ffansatz}), (\ref{hqffansatz}) 
where we distinguish between light flavours 
($D_{u}^{\rm res.}=D_{d}^{\rm res.}=D_{s}^{\rm res.}$), heavy charm and bottom quarks
($D_{c}^{\rm res.}, D_{b}^{\rm res.}$) and gluons ($D_{g}^{\rm res.}$)
and where $D_i^{\rm res.} > 0$ serves as a consistency check.
The residual functions amount to a sufficiently 
small correction not to constrain them any further.
For the same reason the approximation that the hadronic residue 
is half positively, half negatively charged for any parton, i.e.\
$D_i^{{\tiny \Sigma} h^{+,-}}= D_i^{\pi^{+,-}+K^{+,-}}+D_i^{\rm res.}/2$
can be safely used for the inclusive sum of only positively (negatively)
charged hadrons.
Altogether,
the Ans\"{a}tze (\ref{pansatz})-(\ref{kansatz}), (\ref{residual}) 
with the constraints (\ref{parcon}) will be fed into the 
evolution equation (\ref{apff}) to evaluate the cross section 
(\ref{epluseminus}) using the factorization theorem (\ref{fragconv}).
The $\chi^2$-minimization algorithm {\tt MINUIT} \cite{minuit} will be used to obtain
best possible agreement of the outcome with experimental 
data \cite{aleph,sldpk,tpc} over
the range $0.05<z<0.8$. 
Towards low $z$
the TPC data for charged pion production \cite{tpc} at 
$\sqrt{s}=29\ {\rm GeV}$
lie above the corresponding SLD data \cite{sldpk} at $\sqrt{s}=M_Z$ 
($\left. d \sigma^{\rm TPC}\right|_{z\lesssim 0.1} >
\left. d \sigma^{\rm SLD}\right|_{z\lesssim 0.1}$) 
which contradicts the established \cite{bkk4,aleph} QCD evolution
($\left. d \sigma^{\rm QCD}/d s\right|_{z\lesssim 0.1} > 0$)
predicting that $d\sigma$ {\it increases} with $\sqrt{s}$ at low $z$. 
Since the SLD measurements are compatible with the LEP data of e.g.\ ALEPH
\cite{aleph} we exclude TPC charged pion data below $z<0.1$. 
Data on the longitudinal structure function $d\sigma_L^{{\tiny \Sigma} h^\pm}$ 
available
from ALEPH \cite{aleph} will be considered for consistency but
are not weighted in the fit because we found no significant statistical
impact of these data on the fit results. Although one might hope 
$d\sigma_L$ to constrain the gluon fragmentation function because
the gluon enters at `leading' \alps1\ a closer NLO inspection \cite{binne} 
reveals that $d\sigma_L$ is dominated by quark fragmentation over
most of the $z$ range.

\section{Results and Discussion}
\label{results}
\begin{figure}[p]
\vspace*{-.5cm}
\hspace*{-1.cm}
\epsfig{figure=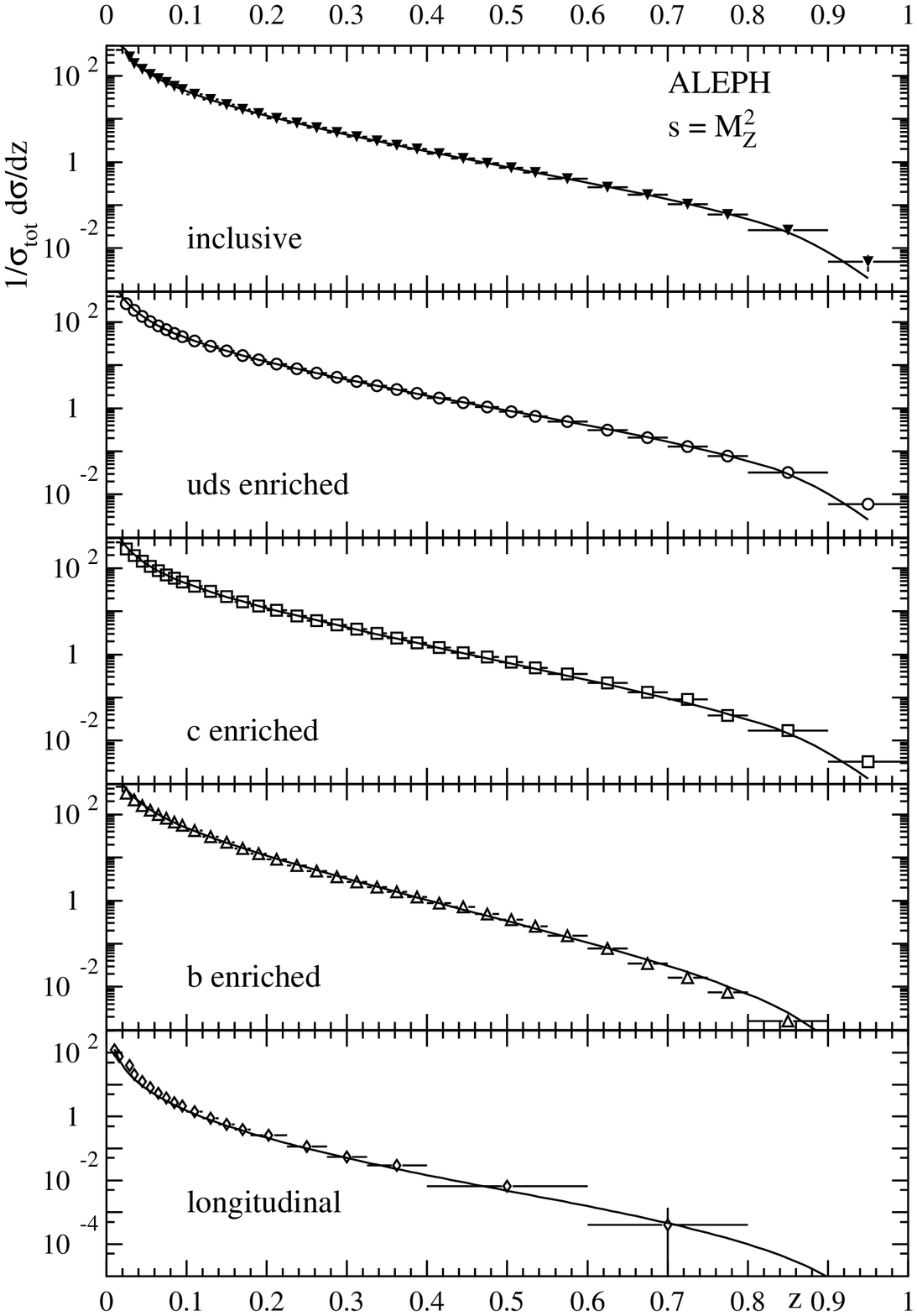,width=16cm}
\vspace*{-2cm}
\caption{ALEPH \cite{aleph} $\sum h^\pm$ inclusive particle spectra, measured at 
the $Z^0$ pole,  and the corresponding
fit results. Details to the individual data samples and curves are given in the text. 
The `longitudinal' set has not been included in the fit.
\label{alephfig}}
\end{figure}
\begin{figure}[t]
\vspace*{-1.5cm}
\hspace*{1.5cm}
\epsfig{figure=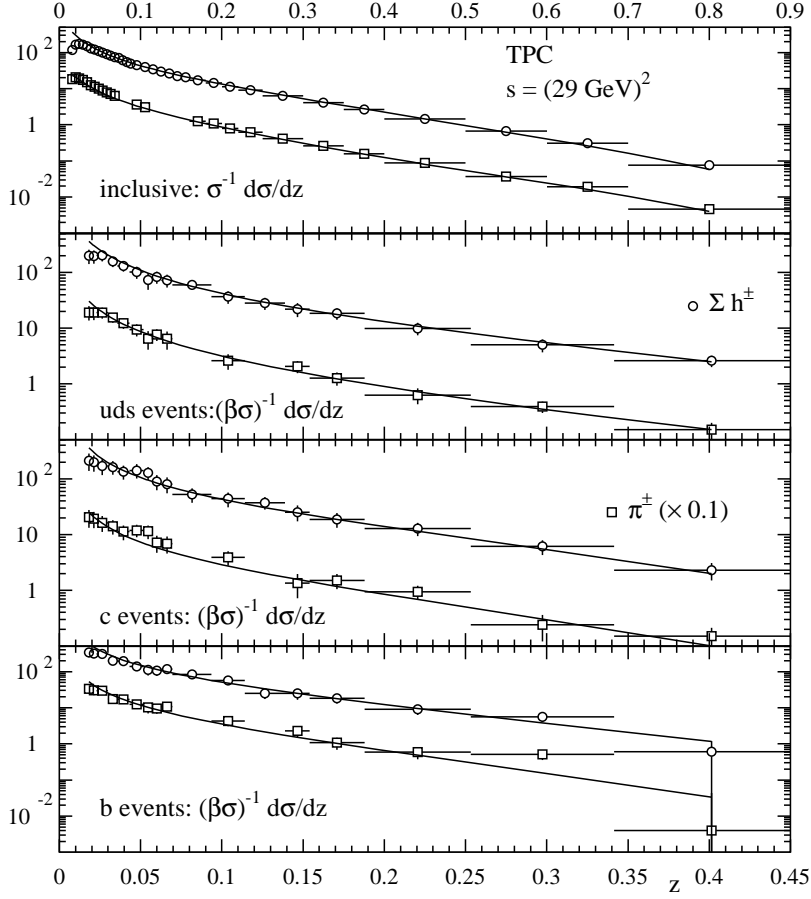,width=12cm}
\vspace*{-3.5cm}
\caption{$\sum h^\pm$ and $\pi^\pm$ inclusive particle spectra
and flavour separated events over the range $0<z<0.9$ (inclusive) and 
$0<z<0.45$ ($\{u,d,s\}$, $c$, and $b$ events; 
$\beta=p_\pi/E_\pi\simeq 1$ for not too small $z$) as
measured at $\sqrt{s}=29\ {\rm GeV}$ by TPC \cite{tpc}.  
The corresponding curves are the fit results. 
Details to the individual data samples and curves are given in the text. 
Also included in the fit were inclusive $K^\pm$ data by TPC which are,
however, not accompanied by flavour tagged samples and are therefore not shown
in the Figure for clearness.
\label{tpcfig}}
\end{figure}
\begin{figure}[t]
\vspace*{-1.cm}
\hspace*{-1.25cm}
\epsfig{figure=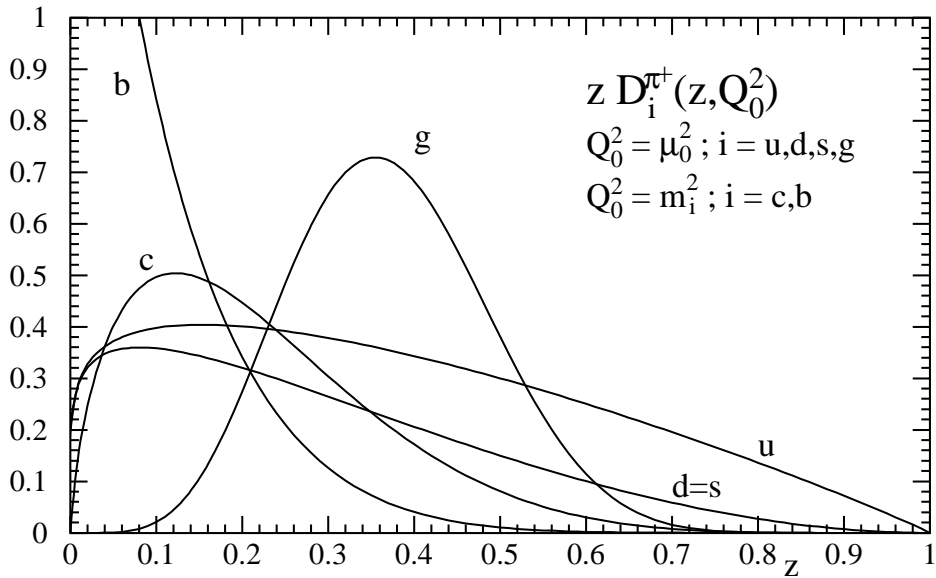,width=16cm}
\vspace*{-2.5cm}
\caption{The input fragmentation functions $D_i^{\pi^+}$ 
as given in Table \ref{fitpars}
at their respective input scales.
\label{input}
}
\end{figure}
\begin{figure}[h]
\vspace*{-1cm}
\hspace*{-1.25cm}
\epsfig{figure=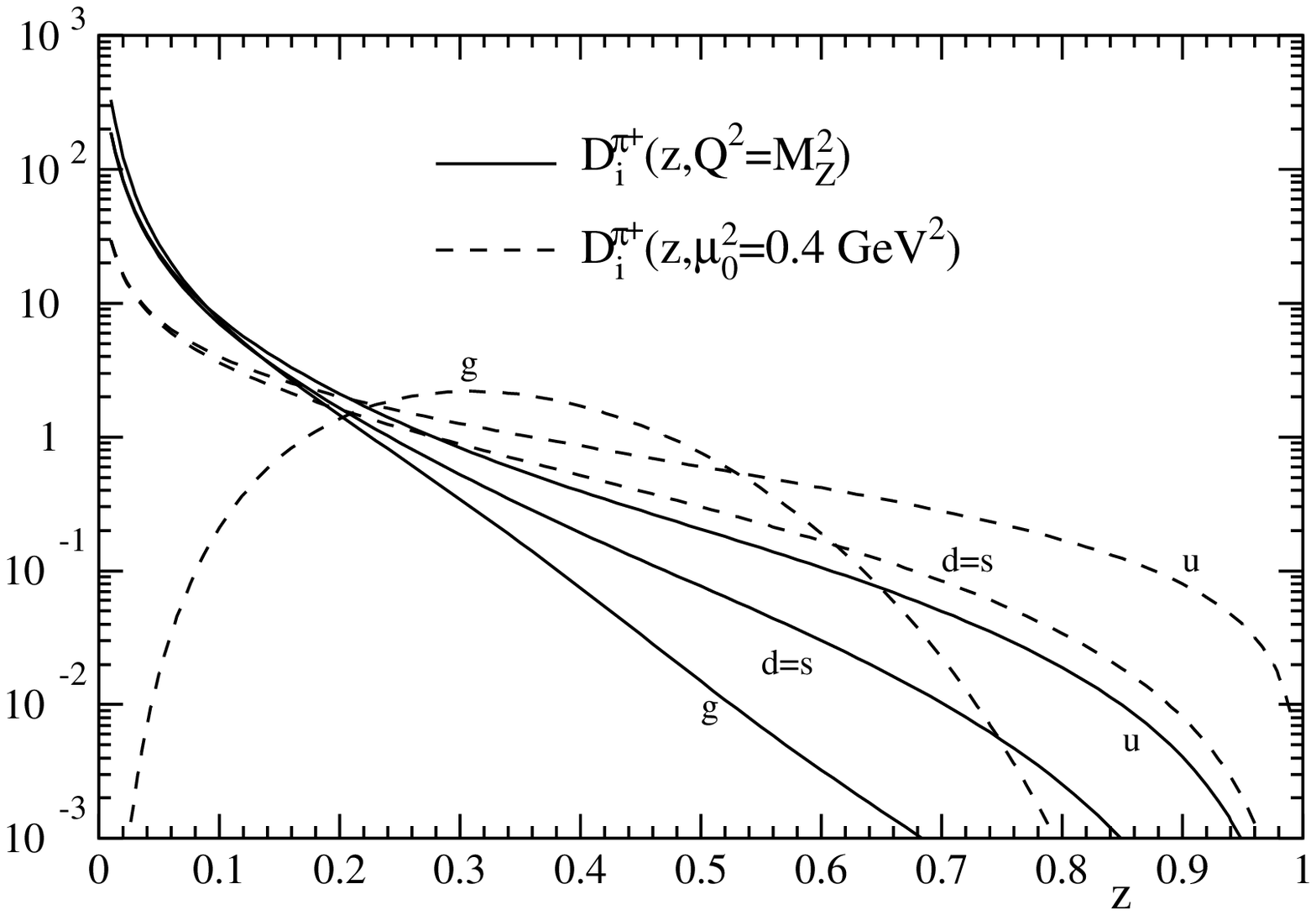,width=16cm}
\vspace*{-1.cm}
\caption{The input for light ($uds$) quarks and gluons of Fig.\ \ref{input}
evolved upward to $Q^2=M_Z^2$. 
\label{ffh100}
}
\end{figure}

\begin{table}[t]
\vspace*{-0.5cm}
\hspace*{2cm}
\begin{tabular}[t]{|c|c|c|}
\hline
$D_i^h(z,Q_0^2)$  & $N_i^h\ z^{\alpha_i^h}\ (1-z)^{\beta_i^h}$ 
&  $D_i^h(n=2,Q_0^2)$          \\  \hline
$D_{u,{\bar d}}^{\pi^+}(z,\mu_{0}^2)$ & $N_u^{\pi^+} z^{-0.829}
(1-z)^{0.949}$ & 0.264 \\ 
$D_{s,{\bar s}}^{\pi^+}(z,\mu_{0}^2)$ & 
$N_u^{\pi^+} z^{-0.829}
(1-z)^{1.949}$ & 0.165\\ 
$D_{g}^{\pi^+}(z,\mu_{0}^2)$ &  
$N_g^{\pi^+} z^{4.374}
(1-z)^{9.778}$ & 0.215\\ 
$D_{c,{\bar c}}^{\pi^+}(z,m_c^2)$ & 
$N_c^{\pi^+} z^{-0.302}
(1-z)^{5.004}$ & 0.166 \\ 
$D_{b,{\bar b}}^{\pi^+}(z,m_b^2)$ & 
$N_b^{\pi^+} z^{-1.075}
(1-z)^{7.220}$ & 0.227\\ \hline
$D_{\bar s}^{K^+}(z,\mu_{0}^2)$ & $N_{\bar s}^{K^+} z^{1.072}
(1-z)^{1.316}$ & 0.148  \\ 
$D_{u}^{K^+}(z,\mu_{0}^2)$ & 
$N_{\bar s}^{K^+} z^{1.072}
(1-z)^{2.316}$ & 0.064 \\
$D_{d,{\bar d}}^{K^+}(z,\mu_{0}^2)$ & 
$N_{\bar s}^{K^+} z^{1.072}
(1-z)^{3.316}$ & 0.033 \\ 
$D_{g}^{K^+}(z,\mu_{0}^2)$ & 
$N_g^{K^+} z^{5.610}
(1-z)^{5.889}$ & 0.048 \\ 
$D_{c,{\bar c}}^{K^+}(z,m_c^2)$ & 
$N_c^{K^+} z^{0.589}
(1-z)^{5.162}$ & 0.074 \\ 
$D_{b,{\bar b}}^{K^+}(z,m_b^2)$ & 
$N_b^{K^+} z^{-0.086}
(1-z)^{7.998} $ & 0.052\\ \hline
$D_{q}^{\rm res.}(z,\mu_{0}^2)$ & $N_q^{\rm res.} 
\ z^{1.006}\ (1-z)^{5.843}$  & 0.043 \\
$D_{g}^{\rm res.}(z,\mu_{0}^2)$ & $ N_g^{\rm res.}
\ z^{6.387}\ (1-z)^{6.435}$ & 0.088\\ 
$D_{c}^{\rm res.}(z,m_c^2)$ & $ N_c^{\rm res.}
\ z^{-1.103}\ (1-z)^{3.917}$ & 0.082 \\ 
$D_{b}^{\rm res.}(z,m_b^2)$ & $ N_b^{\rm res.}
\ z^{-0.605}\ (1-z)^{3.330}$  & 0.113\\ \hline  
\end{tabular}
\caption{\label{fitpars} 
Numerical values for the NLO fit parameters in
Eqs.\ (\ref{pansatz})-(\ref{kansatz}), (\ref{residual}).
The input scales are $\mu_0^2,m_c^2,m_b^2=0.4, 1.96, 20.25 {\rm GeV^2}$
\cite{grv98}. 
The Normalizations $N_i^h$ are determined
by the second moments of the input functions. These moments evolve rather mildly
(typically less than 10\% up to $M_Z^2$)
to higher scales due to the constraint (\ref{parcon}).}
\end{table}
\begin{figure}[t]
\vspace*{-2.cm}
\hspace*{-1.cm}
\epsfig{figure=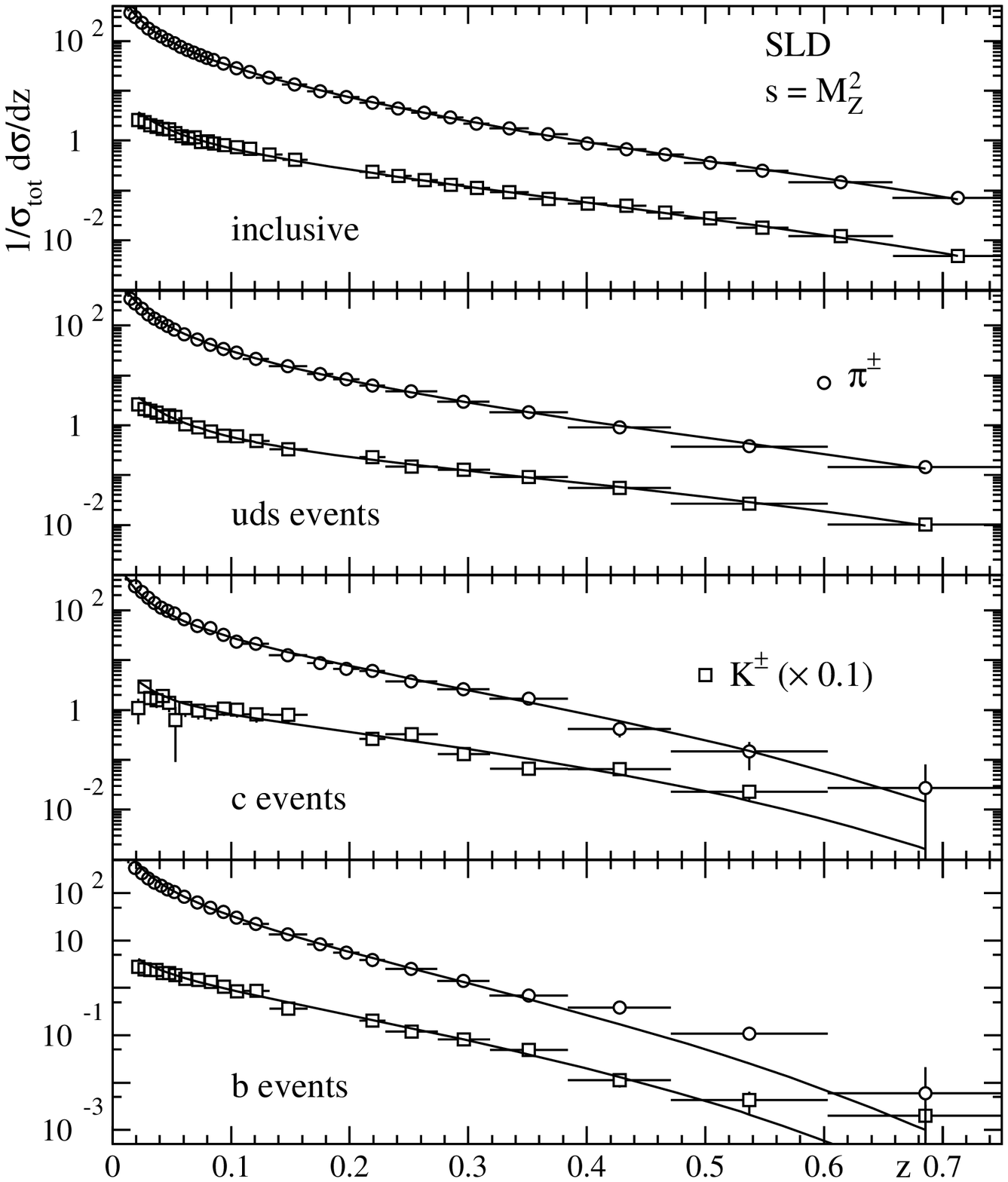,width=15cm}
\vspace*{-4.5cm}
\caption{SLD \cite{sldpk} $\pi^\pm$ and $K^\pm$ inclusive particle spectra,
measured at the $Z^0$ pole, and the 
corresponding fit results. Details to the individual data samples are given in the text.
\label{sldfig}}
\end{figure}
The quality of the fit can be inferred from
Figs.\ \ref{alephfig}, \ref{sldfig}, and \ref{tpcfig}. 
All non-$b$ event samples can be well described within errors, even the
(non-$b$-tagged) ALEPH data where the error is dominated by a normalization
uncertainty of only 1\%. 
As already noted in \cite{aleph,binne} the $b$ quark fragmentation spectrum can
not be described perfectly well using the simple functional form (\ref{hqffansatz})
as input to the evolution. 
The $b$ sets are only reproducible at some 5\% accuracy. 
With the inherent uncertainty due to weak decays we 
consider this a reasonable precision and do not investigate more involved
functional Ans\"{a}tze for the $D_b^h$.
Note that the longitudinal
cross section (lowest curve in Fig.\ \ref{alephfig}), calculated along 
\cite{rijne}, has not been included in the fit. 
We list in Table \ref{fitpars}
the input functions introduced in 
Eqs.\ (\ref{pansatz})-(\ref{kansatz}), (\ref{residual}) 
along with their contributions to the
energy integral (\ref{esum}).
As a representative we show the set $D_i^{\pi^+}$ of input fragmentation 
functions into (positively) charged pions in Fig.\ \ref{input}. 
Evolution effects to higher scales can be inferred for the low input scale
($\mu^2_0=0.4\ {\rm GeV}^2$ \cite{grv98}) functions $D_{i=u,d,s,g}^{\pi^+}$
from Fig.~\ref{ffh100}. They are quite dramatic for the gluon FF but 
less pronounced for the light flavours.
For the higher scale ($m_{c,b}^2$) input functions 
$D_{c,b}^{\pi^+}$, not shown in 
this Figure, the evolution effects are of course 
still weaker as for $D_{u,d,s}^{\pi^+}$.  
The peculiar shape of the input $D_g^{\pi^+}(z,\mu^2_0)$ should be traced
back to starting the evolution at the low input scale 
of \cite{grv98} thereby maximizing the driving force of $D_g^{\pi^+}$.
From pion dominance we would assume a value of
$D_i^{{\tiny \Sigma} h^\pm}(n=2)\lesssim 2/3$ and it is interesting that this
expectation is roughly confirmed. 
Also, estimating the contributions from neutral hadrons from
${\rm SU(2)_f}$ symmetry results in values for the r.h.s.\ of Eq.\ (\ref{esum})
within $[0.9 ; 1.0]$ which 
confirms energy conservation (\ref{conserve}) 
except for $D_b^{{\tiny \Sigma} h}$
which violates  (\ref{conserve}) to about $\sim 10\%$. 
However, as mentioned above, a pure QCD fit to 
bottom fragmentation is 
questionable due to weak 
decay channels which mostly contaminate $b$ jets. 
Furthermore,
a physical interpretation of the second moment integrals 
(\ref{mellin}) over the range $z \epsilon [0;1]$
is delicate and may be  
misleading in general
due to a sizable contribution from the perturbatively 
unstable very low $z$ region,
as we will discuss in more detail in Section \ref{lo}.

\begin{figure}[h]
\vspace*{-0.5cm}
\hspace*{-1.25cm}
\epsfig{figure=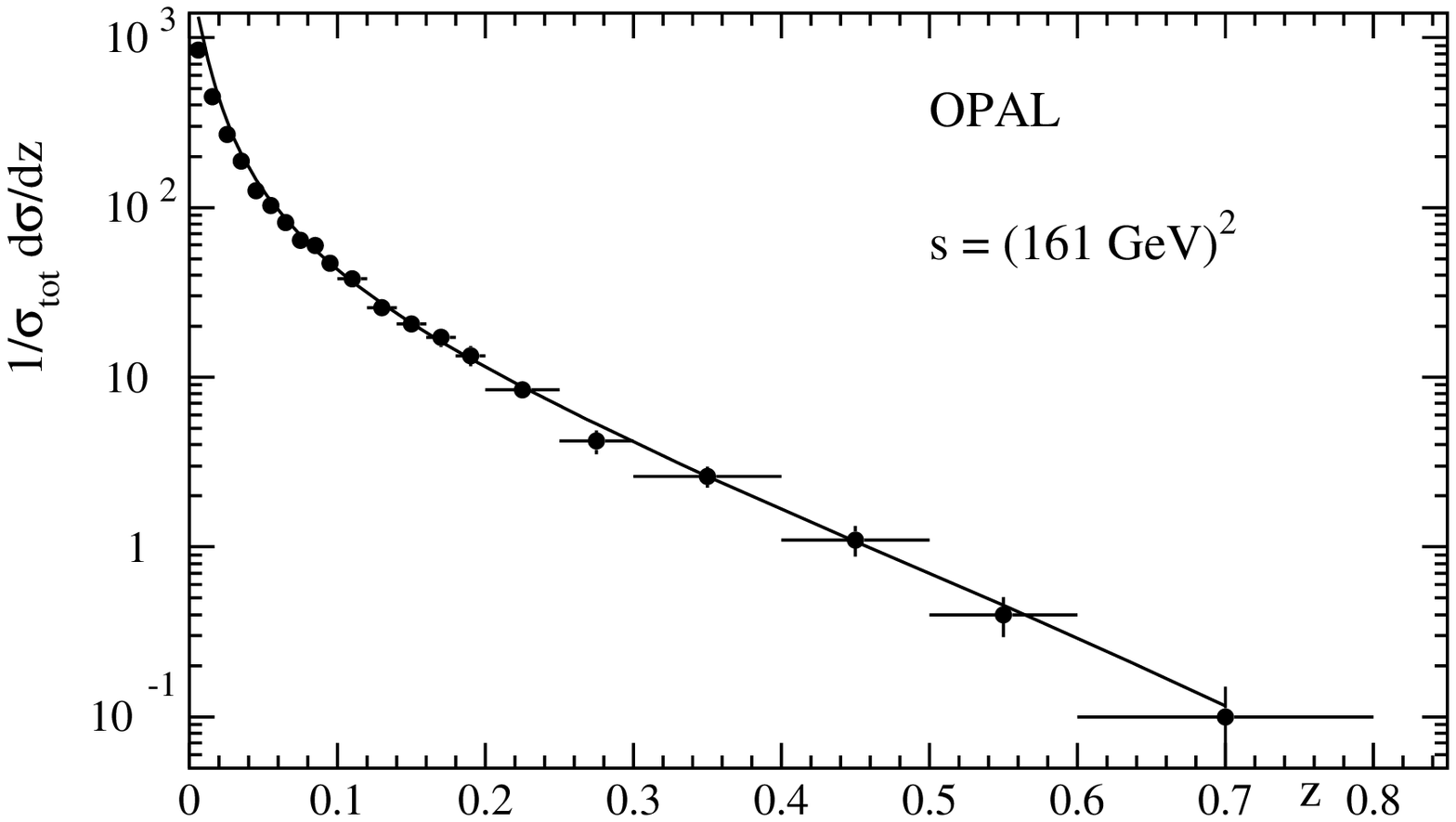,width=16cm}
\vspace*{-2.5cm}
\caption{OPAL inclusive $\sum h^\pm$ particle spectrum measured at a c.m.s.\ energy
of $\sqrt{s}=161\ {\rm GeV}$. The curve is the fit of Fig.\ \ref{alephfig} evolved
upward to that energy. 
\label{opalfig}
}
\end{figure}

\begin{figure}[h]
\vspace*{-1.cm}
\hspace*{-1.25cm}
\epsfig{figure=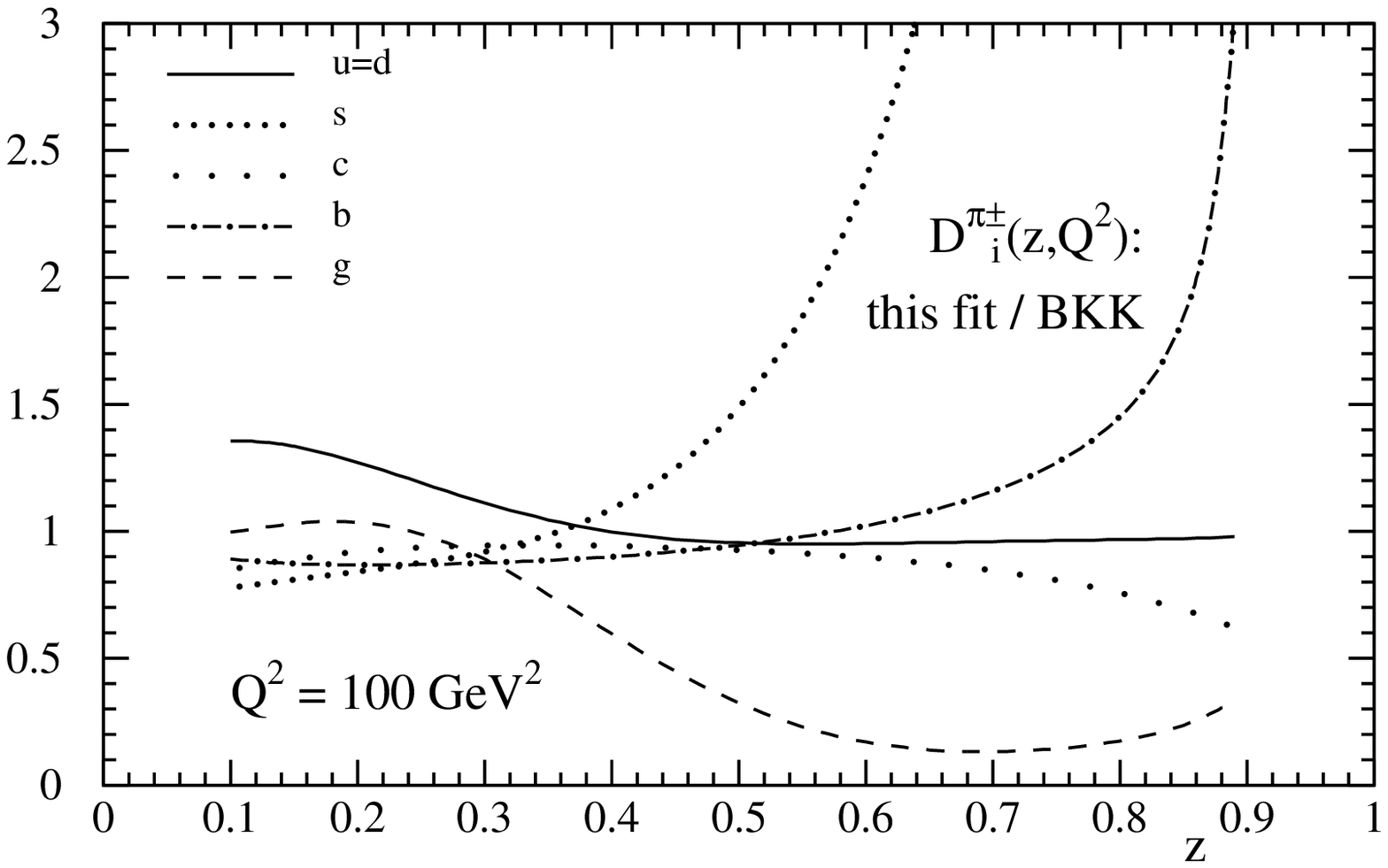,width=16cm}
\vspace*{-2.5cm}
\caption{Ratios of the individual fragmentation functions obtained from
this NLO fit to their analogues in Ref.\ \cite{bkk4}.
\label{pibkk}
}
\end{figure}

\begin{figure}[h]
\vspace*{-0.5cm}
\hspace*{-1.25cm}
\epsfig{figure=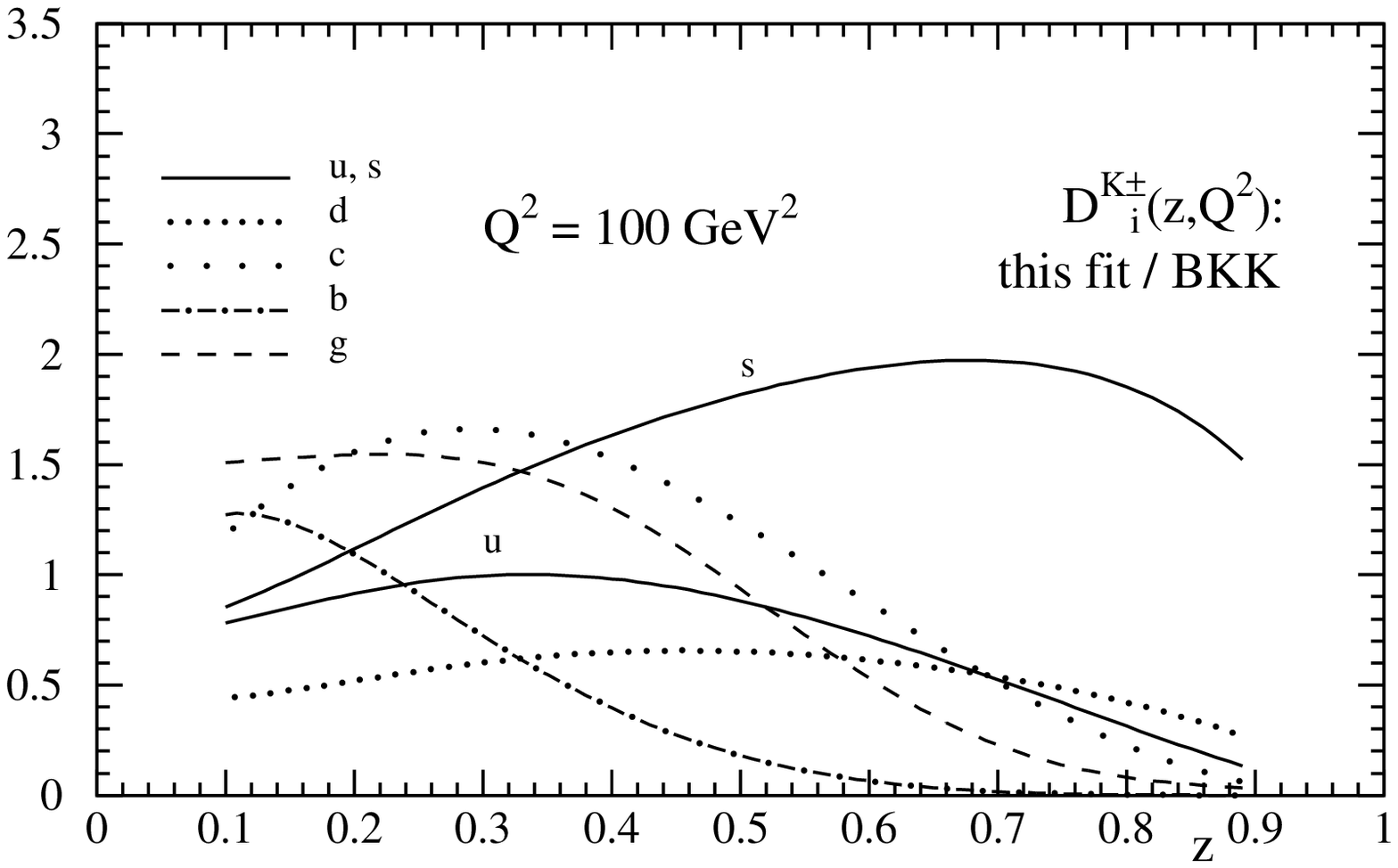,width=16cm}
\vspace*{-2.5cm}
\caption{Same as Fig.\ \ref{pibkk} but for the fragmentation functions into
charged Kaons $K^\pm$.
\label{kbkk}
}
\end{figure}

We have also evolved our $\sum h^\pm$ fit to a c.m.s.\ energy of 
$\sqrt{s}=161\ {\rm GeV}$ and compared the evolution to measurements by 
OPAL \cite{opal161} in Fig.\ \ref{opalfig}. The agreement is convincing albeit
not too surprising because scaling violations from $M_Z\rightarrow 161\ {\rm GeV}$
are rather moderate.  
In Figs.\ \ref{pibkk} and \ref{kbkk} we compare our fitted
fragmentation functions to 
a previous NLO fit by Binnewies, Kniehl and Kramer (BKK) \cite{bkk4}. 
These authors confirm QCD scaling violations within a wider range of data, 
albeit not with the flavour information of Ref.\ \cite{sldpk}
on the individual $\pi^\pm$ and $K^\pm$ spectra. Instead, the flavour separation in
\cite{bkk4} was done by fitting to the flavour tagged ALEPH \cite{aleph}
$\sum h^\pm$ data and assuming $d \sigma^{{\tiny \Sigma} h^\pm}=d \sigma^{(\pi+K)^\pm}+f$ 
where $f$ is a small residual 
from (anti-)protons as measured in \cite{protrat}. 
Besides discrepancies in the barely constrained $D_g^h$,
our procedure of decoupled fits to
$\sum h^\pm$, $\pi^\pm$ and $K^\pm$ data yields quite different results for
the individual flavour fragmentation functions into $\pi^\pm$ and $K^\pm$.
Note that the differences of our fit to BKK decrease stepwise the more 
flavour-inclusive sums of FFs are considered. For the `democratic' FF 
$D_{u+d+s+c+b}^{(\pi+K)^\pm}$ the difference shrinks to 
at most 4\% within $0.1<z<0.8$.
In any case, the light flavour ($uds$) structure is, at present,
arbitrary to some extent
in our fit as well as the one of Ref.\ \cite{bkk4} and differences 
between the two fits of the individual
$D_{i=u,d,s}^{{\tiny \Sigma} h^\pm}$ estimate the present uncertainty on 
these functions.

\subsection{3-Jet Measurements and Gluon Fragmentation; 
Leading Particle Effects}

Complementary to our considerations of one-particle inclusive 
$e^+ e^-$ annihilation data, individual quark and gluon fragmentation 
functions can also be derived from 3-jet topologies
\cite{3jets,opalgjets,delphijets} which can at the leading perturbative 
order be attributed to partonic $q {\bar q} g$ configurations and where
an experimental fragmentation function can be defined as  
\begin{equation}
\label{expfrag}
D_i^h(x_E,\mu^2)\equiv
\frac{1}{N_{\rm tot}}\ \frac{\Delta N_i^h}
{\Delta x_E}; \ \ \ x_E\equiv \frac{E_h}{E_i^{\rm jet}}
\ \\ ;
\end{equation}
i.e.\ as the spectrum in energy $E_h$
scaled to the jet energy $E_i^{\rm jet}$ of some hadron species $h$ 
distributed inside a jet initiated by a parton $i$. 
In general, QCD scale dependence 
from leading logarithms 
can be understood non-covariantly
from the transverse phase space 
volume available for collinear parton emission 
({\it splitting} process)
\begin{equation}
\int_{\mu^2}^{(p_{\perp}^{\max})^2} \frac{d p_{\perp}^2}{p_{\perp}^2} 
\left[P_{ji}^{(0)}(z)\right]_{p_{\perp}=0}
= P_{ji}^{(0)}(z) \ln \left( \frac{p_{\perp}^{\max}}{\mu} \right)^2
\end{equation} 
and a phase space boundary $(p_{\perp}^{\max})^2={\cal{O}}(s)$ leads to 
the common choice $\mu=\sqrt{s}=Q$ for an (one-particle-)inclusive 
phase space. A different $p_{\perp}^{\max}$ is, however, induced in 
a three jet topology from the requirement that parton emission
proceeds into a cone the geometry of
which is defined by the hadrons which are 
grouped together as the jet. The analysis in \cite{delphijets} has 
demonstrated the fragmentation function defined in Eq.\ (\ref{expfrag}) to
undergo scaling violations compatible with LO QCD evolution
in the jet topology scale 
$\mu=E^{\rm jet}\sin (\theta/2)$
where $\theta$ is the angle to the nearest jet in a 3-jet event.        
For the time being, these considerations have to be restricted to a 
LO treatment, since NLO corrections (stemming e.g.\ from a kinematical
configuration where the quark of the $q {\bar q} g$ triple
emits a high $p_{\perp}$ gluon into 
the jet clustered around the anti-quark)  
have to our knowledge not been
formulated yet and we have therefore refrained from including the data 
of \cite{3jets,opalgjets,delphijets} in our fits.  
The gluon fragmentation function $D_g^h$ is, however, barely constrained 
from one-particle-inclusive final state measurements in $e^+e^-$ annihilation
because it enters the cross section in  Eqs.\ (\ref{fragconv}), 
(\ref{epemcoeffs}) only in 
subleading order \alps1\ and where the leading part is factorized into the evolution (\ref{apff})
where $D_g^h$ mixes with the quark singlet fragmentation function
$\sum_q D_q^h$. Therefore, keeping 
the vagueness of this comparison in mind   
we use the 3-jet data of Ref.\ \cite{opalgjets}
to compare our fitted $D_g^{{\tiny \Sigma} h^\pm}(z,\mu^2)$ with, i.e.\ we compare
$D_g^{{\tiny \Sigma} h^\pm}\left(z=x_E,\mu^2=\left<E_g^{\rm jet}\right>^2\simeq 
(40{\rm GeV})^2\right)$ 
with the measured
$\frac{1}{N_{\rm tot}}\ \frac{\Delta N}{\Delta x_E}$ 
in Eq.\ (\ref{expfrag}) for the time
being, with the {\it caveat}
that this cannot give us more than an idea of
the compatibility of our NLO gluon FF with 3-jet measurements.
For illustration we also include the gluon fragmentation 
function of \cite{bkk4} [using the approximation (\ref{hpmapprox})]
and an independent experimental LO determination from DELPHI
\cite{delphijets} in our comparison.  
Hence, the discrepancy between the data
\cite{opalgjets} and the LO QCD fit to an independent 
measurement \cite{delphijets} estimates the accuracy 
to which the 3-jet data \cite{opalgjets} can be identified with a 
QCD gluonic fragmentation function. 
\begin{figure}[h]
\vspace*{-4.cm}
\hspace*{-1.25cm}
\epsfig{figure=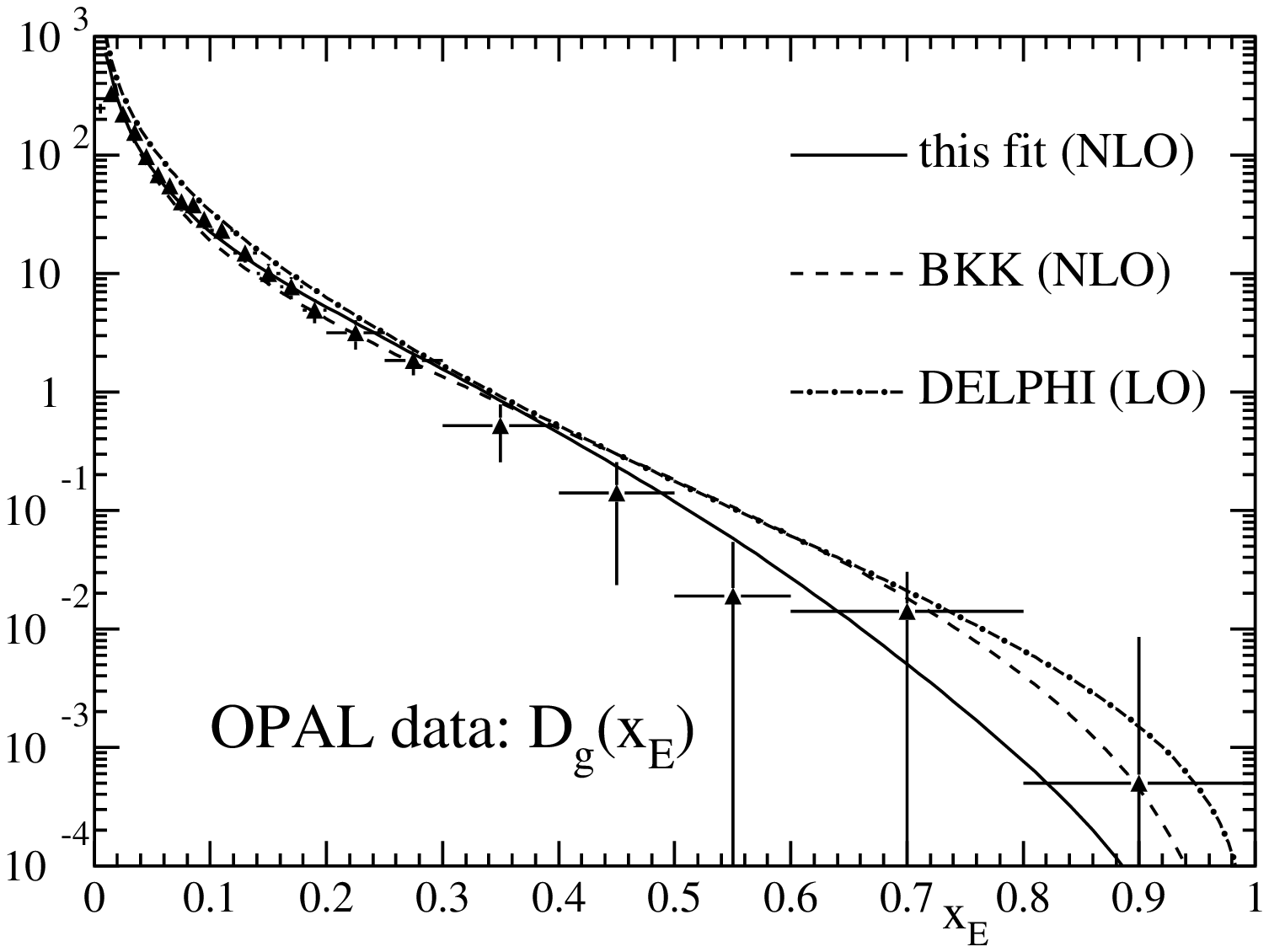,width=16cm}
\vspace*{-3.5cm}
\caption{ The gluon fragmentation function as defined 
experimentally in (\ref{expfrag}) and measured by OPAL \cite{opalgjets}  
compared to the NLO fits of our analysis as well as the one in 
\cite{bkk4}. An experimental LO determination from DELPHI \cite{delphijets}
is also included for comparison.
\label{opaldg}
}
\end{figure}
From Fig.\ \ref{opaldg} 
we judge that the result seems promising and that 
a refinement of the theoretical framework would probably 
contribute to removing the existing ambiguities in the gluon 
fragmentation function.


Similar theoretical limitations as outlined above for 3-jet measurements 
also prevent at present a detailed pQCD analysis of leading particle 
effects \cite{lpeff} in $e^+ e^-$ annihilations which are based experimentally
on phase space restrictions which do not fully match   
a one-particle inclusive QCD framework.
Qualitatively, the results of the leading particle measurements in
\cite{lpeff} are accounted for by Eq.\ (\ref{ffsym6}).

\subsection{Perturbative Stability and Energy Sum Rule}
\label{lo}

\begin{figure}[h]
\vspace*{-4.cm}
\hspace*{-1.25cm}
\epsfig{figure=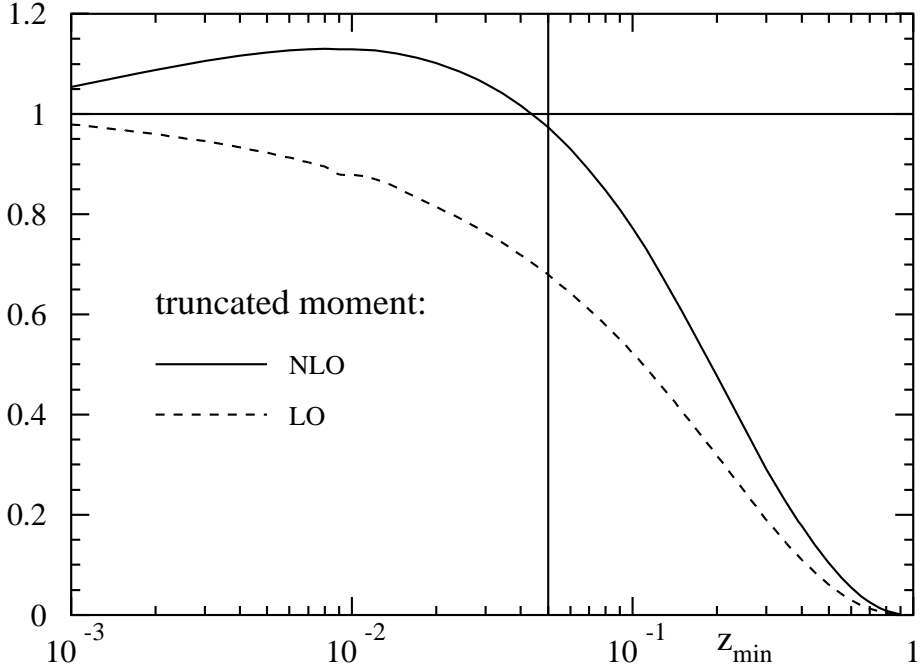,width=16cm}
\vspace*{-3.5cm}
\caption{The truncated second moment of the u-quark contribution 
to charged pion production as defined in Eqs.\ (\ref{flavour}), 
(\ref{truncated}). Shown are results for our NLO and LO fit.
\label{momint}
}
\end{figure}

The full NLO framework of Section \ref{NLOsec} is 
expected
to yield more reliable and less scale-sensitive results 
compared to a LO truncation of the perturbation series
where the known \oalps\ terms of the 
coefficient functions (\ref{epemcoeffs}) are neglected as well as 
the $\beta_1$ contribution to the running of $\alpha_s$
and where the omission of the $P_{ij}^{(1)}$ parts of the 
splitting functions reduces the evolution (\ref{apff}) to
summing only the most dominantly 
leading logs $(\alpha_s/2 \pi \ln Q^2)^n$ for all $n$. 
Still, 
an accompanying LO fit is - besides future 
effective LO applications -
valuable as a test of the perturbative LO$\leftrightarrow$NLO 
stability which is a 
delicate requirement
for perturbative (QCD) approaches to strong interaction phenomena
- especially if the perturbative QCD dynamics is supposed to 
set in at the rather
low input scale in (\ref{ffansatz}) taken from \cite{grv98}.
For infrared {\it un}safe quantities - such as the one-particle-inclusive
spectra considered here - the nonperturbative parameters
have to be redefined at each perturbative order since they replace
new infrared sensitive terms of the factorized perturbation series
at any order. 
We have accordingly performed an accompanying LO fit following exactly
the same lines as the NLO fit described above and resulting in Table
\ref{lofit}.
\begin{table}[t]
\vspace*{-0.5cm}
\hspace*{2cm}
\begin{tabular}[t]{|c|c|c|}
\hline
$D_i^h(z,Q_0^2)$  & $N_i^h\ z^{\alpha_i^h}\ (1-z)^{\beta_i^h}$ 
&  $D_i^h(n=2,Q_0^2)$          \\  \hline
$D_{u,{\bar d}}^{\pi^+}(z,\mu_{0}^2)$ & $N_u^{\pi^+} z^{-0.923}
(1-z)^{0.976}$ & 0.377 \\ 
$D_{s,{\bar s}}^{\pi^+}(z,\mu_{0}^2)$ & 
$N_u^{\pi^+} z^{-0.923}
(1-z)^{1.976}$ & 0.244\\ 
$D_{g}^{\pi^+}(z,\mu_{0}^2)$ &  
$N_g^{\pi^+} z^{5.271}
(1-z)^{8.235}$ & 0.311\\ 
$D_{c,{\bar c}}^{\pi^+}(z,m_c^2)$ & 
$N_c^{\pi^+} z^{-0.818}
(1-z)^{3.461}$ & 0.241 \\ 
$D_{b,{\bar b}}^{\pi^+}(z,m_b^2)$ & 
$N_b^{\pi^+} z^{-1.072}
(1-z)^{6.695}$ & 0.264\\ \hline
$D_{\bar s}^{K^+}(z,\mu_{0}^2)$ & $N_{\bar s}^{K^+} z^{0.617}
(1-z)^{0.744}$ & 0.213  \\ 
$D_{u}^{K^+}(z,\mu_{0}^2)$ & 
$N_{\bar s}^{K^+} z^{0.617}
(1-z)^{1.744}$ & 0.085 \\
$D_{d,{\bar d}}^{K^+}(z,\mu_{0}^2)$ & 
$N_{\bar s}^{K^+} z^{0.617}
(1-z)^{2.744}$ & 0.044 \\ 
$D_{g}^{K^+}(z,\mu_{0}^2)$ & 
$N_g^{K^+} z^{8.132}
(1-z)^{5.776}$ & 0.064 \\ 
$D_{c,{\bar c}}^{K^+}(z,m_c^2)$ & 
$N_c^{K^+} z^{1.419}
(1-z)^{6.171}$ & 0.085 \\ 
$D_{b,{\bar b}}^{K^+}(z,m_b^2)$ & 
$N_b^{K^+} z^{0.191}
(1-z)^{8.934} $ & 0.062\\ \hline
$D_{q}^{\rm res.}(z,\mu_{0}^2)$ & $N_q^{\rm res.} 
\ z^{0.938}\ (1-z)^{7.734}$  & 0.146 \\
$D_{g}^{\rm res.}(z,\mu_{0}^2)$ & $ N_g^{\rm res.}
\ z^{6.150}\ (1-z)^{5.379}$ & 0.003\\ 
$D_{c}^{\rm res.}(z,m_c^2)$ & $ N_c^{\rm res.}
\ z^{-0.636}\ (1-z)^{2.486}$ & 0.084 \\ 
$D_{b}^{\rm res.}(z,m_b^2)$ & $ N_b^{\rm res.}
\ z^{-0.736}\ (1-z)^{3.012}$  & 0.137\\ \hline  
\end{tabular}
\caption{\label{lofit}
Input parameters as
in Table \ref{fitpars} but for our LO fit where 
$\mu_0^2=0.26 {\rm GeV^2}$ \cite{grv98}.  
}
\end{table}
In order not to overload
the logarithmic-scale figures of the preceding sections
with narrow pairs (LO, NLO) of lines and  
instead of repeating the details for the LO fitting procedure we 
rather concentrate in this separate Section on two 
LO$\leftrightarrow$NLO issues worth mentioning.  
\begin{figure}[h]
\vspace*{-4.cm}
\hspace*{-1.25cm}
\epsfig{figure=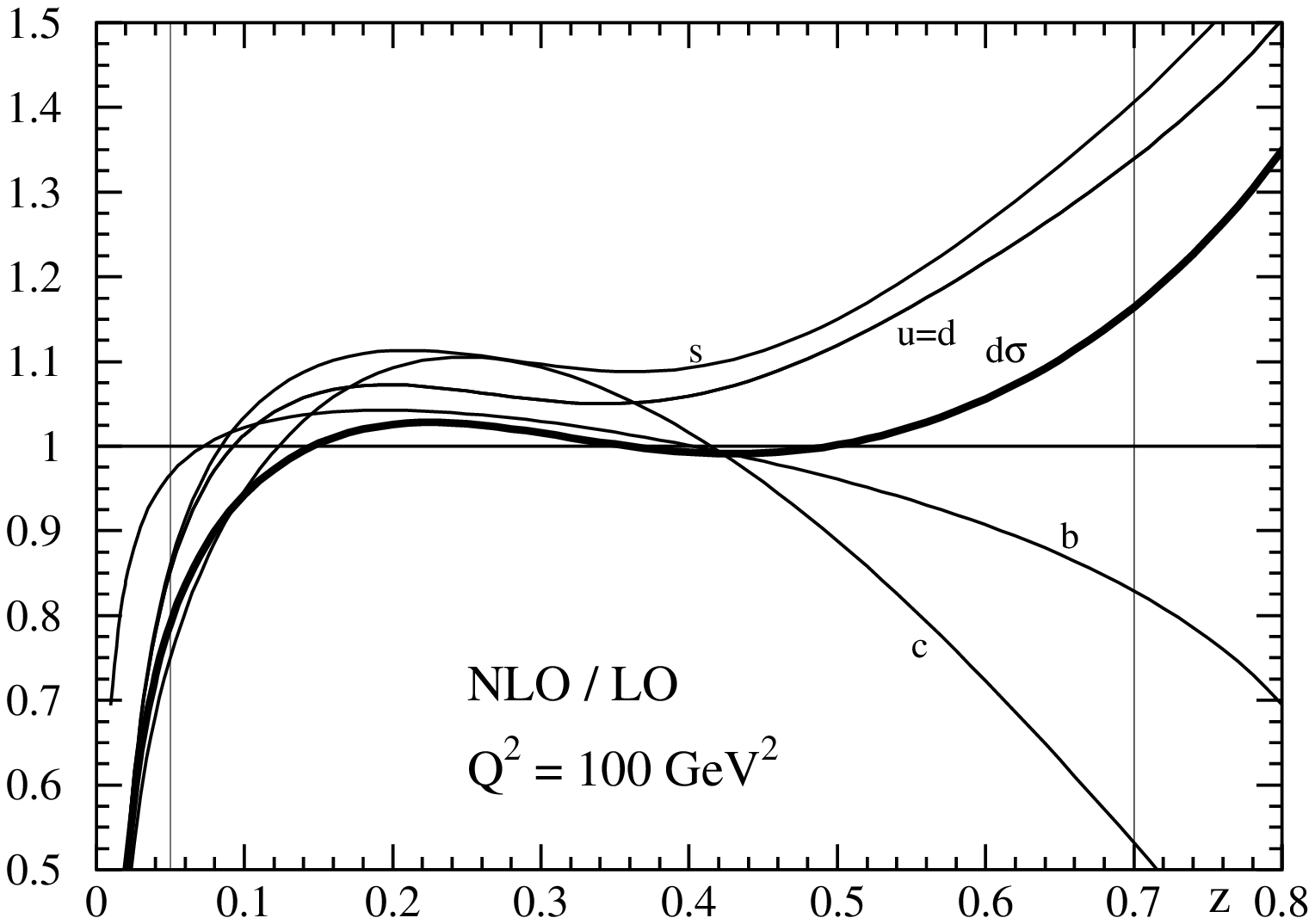,width=16cm}
\vspace*{-3.5cm}
\caption{K-factors for individual flavours and inclusive charged pion
production. The vertical lines indicate the range where the fit is rather
well defined where $z_{\min}$ is set by perturbative stability and 
$z_{\max}$ by experimental statistics.
\label{kfac}
}
\end{figure}
As a representative we will consider our sets of pionic fragmentation 
functions. In Fig.\ \ref{kfac} we show $K$-factors for 
$d \sigma$ in Eq.\ (\ref{epluseminus}), 
$K\equiv d \sigma^h_{\rm NLO}/d \sigma^h_{\rm LO}$, as well as
for each flavour contribution    
\begin{equation}
{\cal D}_q^{\pi^\pm}(z,Q^2)\equiv
\left[ \left( D_{q}^{\pi^\pm}+D_{\bar q}^{\pi^\pm}\right) 
\otimes C_{T+L}^q 
\right] (z,Q^2)
+ \left[ D_g^{\pi^\pm}\otimes {\overline C}_{T+L}^g \right] (z,Q^2) 
\label{flavour}
\end{equation}
where ${\overline C}_{T+L}^g\equiv C_{T+L}^g \times
[ \sigma_0^q /  \sum_{q^{\prime}} \sigma_0^{q^{\prime}} ]$
is the $C_{T+L}^g$ of Eq.\ (\ref{epemcoeffs}) `per flavour' and
where $\otimes$ denotes a convolution integral as in 
Eq.\ (\ref{fragconv}). 
Note that the \alps1\ terms of Eq.\ (\ref{epemcoeffs})
entering Eq.\ (\ref{flavour}) in NLO are neglected in LO
where only the $\sim \delta (1-z)$ term of $C_{T}^q$
contributes. 
The vertical lines in Fig.\ \ref{kfac}
indicate the range 
in $z$ where the fit is rather well determined, 
where $z>0.05$ excludes the perturbatively 
unstable \cite{fsv} low $z$ region and where for $z\gtrsim 0.7$ 
statistics drop rather low (only one data point for $z>0.7$).
Hence, the spread of the curves towards large $z$ 
should not be taken as a perturbative instablitiy but attributed 
mainly to the decreasing experimental statistics which do not 
determine the FFs very well \cite{aurenche} 
in the $z\rightarrow 1$ hard fragmentation limit
and where 
poorly defined $\chi^2$ minima may easily fake perturbative
instability in decoupled LO$\leftrightarrow$NLO fits.
It is therefore reassuring to observe that the 
flavour-inclusive $d\sigma$ has a $K$-factor
closer to one than the
individual flavour-contributions where the latter have 
poorer experimental statistics and larger systematic
errors than the former. 
On the other hand, the unstable $z\rightarrow 0$
behaviour is unaffected by the choice of input
parameters and can be traced back to including/omitting 
the NLO pieces $P_{ij}^{(1)}$ of the splitting functions in the
NLO/LO evolution. Indeed, our choice $z_{\min}=0.05$ appears
already to be on the edge of perturbative reliability.
The far steeper LO evolution has profound impact on
the second moments in Table \ref{lofit} which violate
the energy sum rule (\ref{conserve}). We demonstrate this point
in Fig.\ \ref{momint} where we show the truncated moment
\begin{equation}
\label{truncated}
\frac{ \int_{z_{\min}}^1 d z\ z\ {\cal D}_u^{\pi^+}(z,Q^2)}
{{\cal D}_u^{\pi^+}(n=2,Q^2)}
\end{equation}
in LO and NLO. It can be seen that the full second moment receives 
in LO a sizable contribution from the unstable $z<0.05$ region
where the evolution generates a too steep spectrum resulting in
large second moments violating (\ref{conserve}). On the other hand,
in NLO the second moment, truncated at $z_{\min}=0.05$, is rather close 
to its {\it un}truncated value leading to NLO moments that compare rather 
well with (\ref{conserve}). One must note, however, that the choice
 $z_{\min}=0.05$ is rather arbitrary and that the region 
$z\epsilon [0.0 ; 0.05]$ does not contribute 
much in NLO partly because 
${\cal D}_q^{\pi^\pm}(z,Q^2)$ turns {\it un}physically
{\it negative} at very small $z$
due to the timelike NLO evolution \cite{fsv}. 
This observation renders the energy sum
rule (\ref{conserve}) a delicate 
concept for perturbative QCD FFs and we believe it should not be 
considered within this theoretical framework
unless the $z\rightarrow 0$ behaviour of FFs is under better 
control. 

\section{Conclusions}

Within the radiative parton model \cite{grv90,grv92,grv93,grv94,grv98}
we have fitted parton fragmentation functions to identified
hadron ($\pi^\pm$, $K^\pm$) and inclusive charged particle spectra measured
at the $Z^0$ pole \cite{aleph,sldpk}. 
Scaling violations were properly taken into account by
simultaneously fitting lower energy TPC data. 
Special attention was paid to the flavour structure of the parton FFs
where a collinearly resummed RGE
formalism was argued to be adequate
for the treatment of heavy quark contributions.
In the light quark sector we made power law assumptions which qualitatively
establish a physical hierarchy among the FFs which is 
guided by the ideas of
valence enhancement and strangeness suppression and  
which is observed in leading particle measurements. 
It would be desirable to interpret these measurements - as well as
gluon-jet measurements from 3-jet topologies - 
in more quantitative detail within the language 
of NLO QCD fragmentation functions. 
A corresponding framework has, however, to our knowledge not been 
developed yet. 
Despite the high precision data available, the {\it individual} 
$D_{i=u,d,s,g,c,b}^{{\tiny \Sigma} h^\pm,\pi^\pm,K^\pm}$ are 
therefore still rather uncertain. 
To estimate the present theoretical uncertainty we compared
our fit to the one of \cite{bkk4} and found sizable deviations.
An  inclusive sum over the distinct flavours
is, however,  rather reliably determined for not too large $z$.
The missing experimental information at large $z$ may fake 
perturbative instability in independent LO$\leftrightarrow$NLO
fits manifesting, however, just the fact that the 
FFs are still unknown in the $z\rightarrow 1$ hard 
fragmentation limit.  
We also considered the contributions of the individual 
FFs to the energy integral in (\ref{esum})
and found reasonable values in NLO while the steeper 
LO evolution requires too large moments. The instability
of the timelike evolution at low $z$ makes, however,  
energy conservation a concept which may hardly be useful 
within perturbative QCD even for NLO FFs.

A {\tt FORTRAN} package of the LO and NLO $D_i^h$-functions will be 
obtainable upon request.

%
\section*{Acknowledgements}
We thank E.\ Reya and M.\ Gl\"{u}ck for suggestions
and many instructive discussions,
I.\ Schienbein for helpful discussions and a careful reading, and 
W.\ Bernreuther, A.\ Brandenburg and P.\ Uwer for helpful correspondences. 
We are thankful to W.\ Hofmann for making Tables of the flavour tagged
TPC data accessible to us.
The work has been supported in part by the `Bundesministerium f\"{u}r Bildung, 
Wissenschaft, Forschung und Technologie', Bonn. 

\newpage
\topmargin 0.0cm
\begin{appendix}

\section*{Appendix:\\ 
\msbar\ Coefficient Functions for One Hadron Inclusive $e^+e^-$
Annihilation}

\label{epemapp}

The \alps1\ coefficient functions for One Hadron Inclusive $e^+e^-$
Annihilation as introduced in Section \ref{epemchap} read \cite{aemp,fupe,nasweb}:
\begin{eqnarray} \nonumber
c_T^q\left(\zeta,\frac{Q^2}{\mu^2_F}\right) &=& (1+\zeta^2)\left[\frac{\ln (1-\zeta)}
{1-\zeta}\right]_+ - \frac{3}{2} \left[\frac{1}{1-\zeta}\right]_+
+2 \frac{1+\zeta^2}{1-\zeta}\ln \zeta \\ \nonumber
&+&\frac{3}{2} (1-\zeta)+(\frac{2}{3}\pi^2-\frac{9}{2})\delta(1-\zeta)
+\ln\frac{Q^2}{\mu^2_F}\left[\frac{1+\zeta^2}{1-\zeta}\right]_+ \ \ \ , 
\\ \nonumber 
c_T^g\left(\zeta,\frac{Q^2}{\mu^2_F}\right) &=& 2 \frac{1+(1-\zeta)^2}{\zeta}
\left(\ln(1-\zeta)+2\ln\zeta+\ln\frac{Q^2}{\mu^2_F}\right)-4\frac{1-\zeta}{\zeta}\ \ \ ,
\\ \nonumber  c_L^q\left(\zeta,\frac{Q^2}{\mu^2_F}\right) &=& 1\ \ \  ,
\\ \label{cqgs} c_L^g\left(\zeta,\frac{Q^2}{\mu^2_F}\right) &=&
4 \frac{1-\zeta}{\zeta}\ \ \  . \label{ctlqg}
\end{eqnarray}
The parton model, i.e.\ \alp\, electroweak cross section 
$\sigma_0^q$ for producing a $q{\bar q}$ pair in $e^+e^-$ annihilation are 
given by 
\begin{equation}
\label{sig0}
\sigma_0^q(s) = \frac{4 \pi \alpha^2}{s}\ \left[
e_q^2+2 e_q v_e v_f \rho_1(s)+(v_e^2+a_e^2)(v_q^2+a_q^2) \rho_2(s)\right]
\end{equation}
with the QED fine structure constant $\alpha$ and where 
\begin{eqnarray} \nonumber
\rho_1(s) &=& \frac{1}{4\sin^2\theta_W\cos^2\theta_W} \frac{s(M_Z^2-s)}
{(M_Z^2-s)^2+M_Z^2\Gamma_Z^2}\ \ \ , \\
\rho_2(s) &=& \left(\frac{1}{4\sin^2\theta_W\cos^2\theta_W}\right)^2 \frac{s^2}
{(M_Z^2-s)^2+M_Z^2\Gamma_Z^2}\ \ \ , 
\end{eqnarray}
and the electric charges $e_i$ and electroweak vector ($v_i$) and axial ($a_i$) 
couplings are listed in Table \ref{couptab} according to their standard model values:
\begin{eqnarray} \nonumber
v_i &=& T_{3,i} - 2 e_i \sin^2\theta_W \\
a_i &=& T_{3,i}\ \ \ , 
\end{eqnarray}
where ${\vec T}$ is the weak isospin and $\theta_W$ the Weinberg angle. 
\noindent
\begin{table}
\begin{tabular*}{\textwidth}[t]{@{~}l@{\extracolsep\fill}cccc}
\hline\hline
particle $i$  & $e_i$          & $v_i$                                    & $a_i$  \\  \hline
$e^-$         & $-1$           & $-\frac{1}{2}+2\sin^2\theta_W$           & $-\frac{1}{2}$ \\
up-type quark       & $+\frac{2}{3}$  & $+\frac{1}{2}-\frac{4}{3}\sin^2\theta_W$  & $+\frac{1}{2}$  \\
down-type quark    & $-\frac{1}{3}$ & $-\frac{1}{2}+\frac{2}{3}\sin^2\theta_W$ & $-\frac{1}{2}$\\ 
\hline\hline \\
\end{tabular*}
\caption{\label{couptab}
Electromagnetic and electroweak couplings entering Eq.\ (\ref{sig0})}
\end{table}

The Mellin transforms of the $c_{T,L}^{q,g}$ in 
Eq.\ (\ref{ctlqg}) read \cite{grvff}
\begin{eqnarray}\nonumber
c_T^q\left(n,\frac{Q^2}{\mu^2_F}\right)
&=& 5{\rm S}_2(n)+{\rm S}_1^2(n)+{\rm S}_1(n)\left[\frac{3}{2}-\frac{1}{n(n+1)}\right]
-\frac{2}{n^2}+\frac{3}{(n+1)^2}
\\ \nonumber
&-& \frac{3}{2}\frac{1}{n+1}-\frac{9}{2} +\left[\frac{1}{n(n+1)}-2{\rm S}_1(n)+
\frac{3}{2}\right]
\ln\frac{Q^2}{\mu^2_F} 
\\ \nonumber
c_T^g\left(n,\frac{Q^2}{\mu^2_F}\right)
&=& 2\left[-{\rm S}_1(n)\frac{n^2+n+2}{(n-1)n(n+1)}-\frac{4}{(n-1)^2}
+\frac{4}{n^2}-\frac{3}{(n+1)^2}\right] 
\\ \nonumber
&+& 2\frac{n^2+n+2}{n(n^2-1)}\ln\frac{Q^2}{\mu^2_F}
\\ \nonumber
c_L^q\left(n,\frac{Q^2}{\mu^2_F}\right)
&=& \frac{1}{n}\\ 
c_L^g\left(\zeta,\frac{Q^2}{\mu^2_F}\right)
&=& \frac{4}{(n-1)n}\ \ \ ,
\end{eqnarray}
where the sums
\begin{equation}
\label{sks}
{\rm S}_k(n)\equiv\sum_{j=1}^n\frac{1}{j^k}
\end{equation} 
have to be analytically continued \cite{grv92,grvg} to the complex $n$ plane
\begin{eqnarray} \nonumber
{\rm S}_1(n) &=& \gamma_E+\psi(n+1),\ \gamma_E=0.577216 \\  
{\rm S}_2(n) &=& \zeta (2)-\psi^\prime(n+1),\ \ \ \zeta (2)=\frac{\pi^2}{6} 
\label{s12complex} 
\end{eqnarray}
with the help of logarithmic derivatives of the $\Gamma$ function
$\psi^{(k)}(n)\equiv d^{(k+1)}\ln \Gamma(n)/dn^{k+1}$.
Analogously to the \alps1\ contribution $F_L$ to the deep inelastic electron
nucleon cross section, the \alps1\ contribution from longitudinally polarized 
virtual bosons to the fragmentation spectrum in $e^+e^-$ annihilations is 
scheme independent and finite. The $c_L^{q,g}$ do therefore {\it not} depend
on the factorization scale $\mu^2_F$, contrary to the $c_T^{q,g}$ which are
infrared safe only after factorization of the collinear singularities.  

\end{appendix}


\begin{thebibliography}{99}
\bibitem{forthcoming}
Research to be pursued by the author.
\end{thebibliography}
\end{document}